\documentclass{emulateapj}
\usepackage{natbib}
\usepackage{xcolor}
\usepackage{float}
\usepackage{epsfig}
\usepackage{graphicx}
\usepackage{graphics}
\usepackage[latin1]{inputenc}
\usepackage{latexsym}
\slugcomment{ApJ, in press}

\newcommand{\nd}{\multicolumn{1}{c}{$\dots$}}

\newcommand{\ho}{H$_0$}

\newcommand{\ncep}{2200}
\newcommand{\snia}{SNe~Ia}
\newcommand{\snh}{SN~Ia hosts}
\newcommand{\V}{\it F555W}
\newcommand{\I}{\it F814W}
\newcommand{\W}{\it F350LP}
\newcommand{\irh}{\it F160W}
\newcommand{\mgal}{M101}
\newcommand{\fgal}{N1015}
\newcommand{\xgal}{N1309}
\newcommand{\kgal}{N1365}
\newcommand{\ggal}{N1448}
\newcommand{\egal}{N2442}
\newcommand{\ygal}{N3021}
\newcommand{\zgal}{N3370}
\newcommand{\pgal}{N3447}
\newcommand{\qgal}{N3972}
\newcommand{\rgal}{N3982}
\newcommand{\agal}{N4038}
\newcommand{\ngal}{N4258}
\newcommand{\wgal}{N4424}
\newcommand{\lgal}{N4536}
\newcommand{\bgal}{N4639}
\newcommand{\tgal}{N5584}
\newcommand{\ogal}{N5917}
\newcommand{\dgal}{N7250}
\newcommand{\ugal}{U9391}

\shorttitle{Cepheids in 19 SNe Ia hosts and N4258 with {\it HST}}
\shortauthors{Hoffmann et al.}

\begin{document}

\title{Optical Identification of Cepheids in 19 Host Galaxies of Type Ia Supernovae\\and NGC 4258 with the {\it Hubble Space Telescope}\altaffilmark{1}}

\author{Samantha L.~Hoffmann\altaffilmark{2}, Lucas M.~Macri\altaffilmark{2}, Adam G.~Riess\altaffilmark{3,4}, Wenlong Yuan\altaffilmark{2}, Stefano Casertano\altaffilmark{4},\\Ryan J.~Foley\altaffilmark{5,6,7}, Alexei V.~Filippenko\altaffilmark{8}, Brad E.~Tucker\altaffilmark{8,9}, Ryan Chornock\altaffilmark{10}, Jeffrey M.~Silverman\altaffilmark{11}, Douglas L.~Welch\altaffilmark{12}, Ariel Goobar\altaffilmark{13}, and Rahman Amanullah\altaffilmark{13}}

\altaffiltext{1}{\footnotesize Based on observations with the NASA/ESA {\it Hubble Space Telescope}, obtained at the Space Telescope Science Institute, which is operated by AURA, Inc., under NASA contract NAS 5-26555.}
\altaffiltext{2}{\footnotesize George P.~and Cynthia W.~Mitchell Institute for Fundamental Physics \& Astronomy, Dept.~of Physics \& Astronomy, Texas A\&M Univ., College Station, TX, USA}
\altaffiltext{3}{\footnotesize Dept.~of Physics \& Astronomy, Johns Hopkins Univ., Baltimore, MD, USA}
\altaffiltext{4}{\footnotesize Space Telescope Science Institute, Baltimore, MD, USA}
\altaffiltext{5}{\footnotesize Dept.~of Astronomy \& Astrophysics, Univ.~of California, Santa Cruz, CA, USA}
\altaffiltext{6}{\footnotesize Dept.~of Astronomy, Univ.~of Illinois at Urbana-Champaign, Urbana, IL, USA}
\altaffiltext{7}{\footnotesize Dept.~of Physics, Univ.~of Illinois at Urbana-Champaign, Urbana, IL, USA}
\altaffiltext{8}{\footnotesize Dept.~of Astronomy, Univ.~of California, Berkeley, CA, USA}
\altaffiltext{9}{\footnotesize The Research School of Astronomy \& Astrophysics, Australian National Univ., Mount Stromlo Observatory, Weston Creek, ACT, Australia}
\altaffiltext{10}{\footnotesize Astrophysical Institute, Dept.~of Physics \& Astronomy, Ohio Univ., Athens, OH, USA}
\altaffiltext{11}{\footnotesize Dept.~of Astronomy, Univ.~of Texas, Austin, TX, USA}
\altaffiltext{12}{\footnotesize Dept.~of Physics \& Astronomy, McMaster Univ., Hamilton, ON, Canada}
\altaffiltext{13}{\footnotesize The Oskar Klein Centre, Physics Dept., Stockholm Univ., Stockholm, Sweden}

\begin{abstract}
We present results of an optical search for Cepheid variable stars using the {\it Hubble Space Telescope (HST)} in 19 hosts of Type Ia supernovae (SNe~Ia) and the maser-host galaxy NGC 4258, conducted as part of the SH0ES project (Supernovae and {\ho} for the Equation of State of dark energy). The targets include 9 newly imaged {\snh} using a novel strategy based on a long-pass filter that minimizes the number of {\it HST} orbits required to detect and accurately determine Cepheid properties. We carried out a homogeneous reduction and analysis of all observations, including new universal variability searches in all {\snh}, that yielded a total of {\ncep} variables with well-defined selection criteria, the largest such sample identified outside the Local Group. These objects are used in a companion paper to determine the local value of {\ho} with a total uncertainty of 2.4\%.
\end{abstract}


\section{Introduction \label{sec:intro}}

The Cepheid period-luminosity relation (hereafter, PLR) or ``Leavitt Law'' \citep{leavitt12} is one of the most widely used primary distance indicators and has played a central role in many efforts to determine the local expansion rate of the Universe or Hubble constant \citep[\ho;][]{hubble29}. Six decades' worth of efforts on the extragalactic distance scale \citep[summarized in the reviews by][]{madore91,jacoby92} led to $\sigma\!\approx\!10\%$ determinations of this key cosmological parameter  by \citet{freedman01} and \citet{sandage06} using the {\it Hubble Space Telescope (HST)}. The discovery of the acceleration of cosmic expansion \citep{riess98,perlmutter99} motivated the continued development of increasingly more robust and precise distance ladders to better constrain the nature of dark energy. Building on the discovery of a large sample of Cepheids in NGC 4258 \citep[N4258;][hereafter M06]{macri06} and the promising geometric distance to this galaxy \citep{herrnstein99}, the SH0ES project (Supernovae and H$_0$ for the Equation of State of dark energy) focused on reducing sources of systematic uncertainty that yielded $\sigma(\textrm{H}_0)=5$\%, 3.3\%, and 2.4\% \citep[][hereafter R09a, R11, and R16, respectively]{riess09a,riess11,riess16}.

The most recent of these determinations benefits from many improvements to the distance scale over the past decade, including but not limited to high signal-to-noise ratio (S/N) parallaxes to Milky Way Cepheids \citep{benedict07,riess14,casertano16}, larger samples of Cepheids in the Large Magellanic Cloud (LMC) with homogeneous optical and near-infrared light curves \citep{soszynski08,macri15}, and robust distances to the LMC \citep{pietrzynski13} and \ngal\ \citep{humphreys13}. R16 ties these improvements on the ``first rung'' of the ladder to a sample of 281 Type Ia supernovae ({\snia}) in the Hubble flow through Cepheid-based distances to 19 host galaxies of ``ideal'' {\snia}. The aim of this publication is to present the details of the optical observations, data reduction and analysis, and selection of the Cepheid variables in these {\snh} and the anchor {\ngal}. Near-infrared follow-up observations of these Cepheids are presented in our companion paper (R16). 

The rest of the paper is organized as follows. \S\ref{sec:obsdata} describes the {\it HST} observations and data reduction. Details of the point-spread function (PSF) photometry and calibration steps are given in \S\ref{sec:psfphot}. In \S\ref{sec:idceph} we discuss the Cepheid search and selection criteria, and in \S\ref{sec:results} we address systematic corrections. Our results are summarized in \S\ref{sec:sum}.

\vfill\pagebreak\newpage 

\section{Observations and Data Reduction} \label{sec:obsdata}

{\it HST} observations of Cepheid variables span more than two decades, highlighting the relevance of this topic for the initial development and subsequent mission of the observatory. The earliest Cepheid observations we analyzed were obtained with the Wide Field and Planetary Camera 2 (WFPC2) as part of the initial efforts to measure {\ho} with {\it HST} \citep{freedman01,sandage06} and were later used by \citet{freedman12} to reach beyond the LMC for the Carnegie Hubble Project. The famous chevron-shaped field of view of this instrument consists of three quadrants spanning $80\arcsec$ on a side, sampled at $0\farcs1$~pix$^{-1}$, with the remaining quadrant covering $37\arcsec$~on a side at a finer scale of $0\farcs046$~pix$^{-1}$ \citep{mcmaster08}. Given the overall poorer sampling of the PSF and lower throughput of this camera relative to more modern instruments, we only used these images for time-series information and relied on subsequent observations to generate input star lists and to calibrate the photometry. 

\begin{figure*}
\begin{center}
\includegraphics[width=\textwidth]{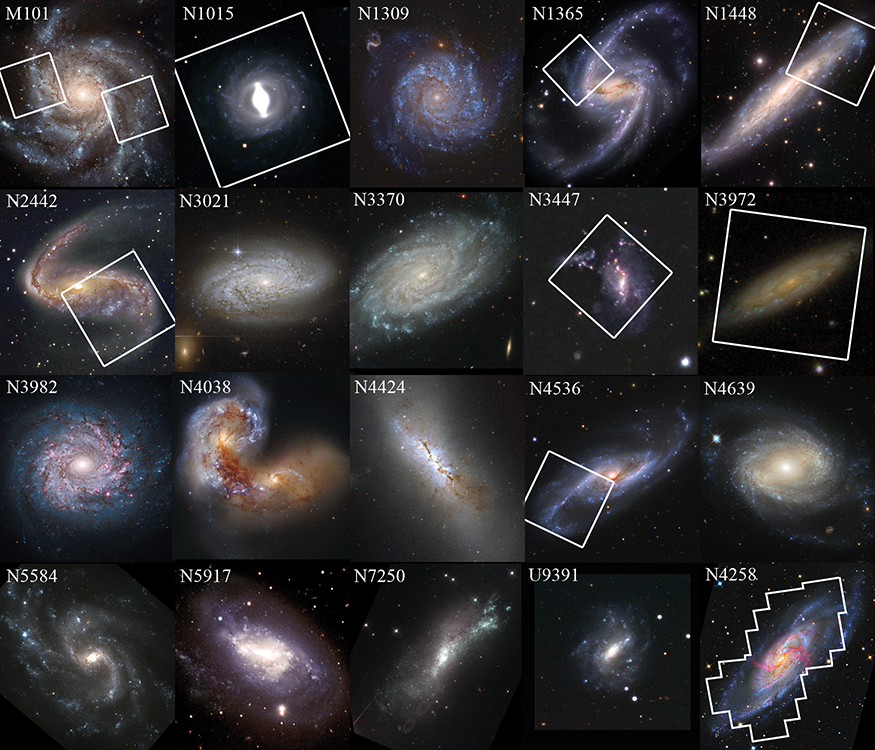}
\caption{Color images of the galaxies studied in this paper. The {\it HST} ACS or WFC3 fields of view are outlined only in those cases where the size of the image is significantly larger.}
\label{fig:galaxies}
\end{center}
\end{figure*}

We also reanalyzed observations obtained in previous phases of our project (\citealt{riess09b}, hereafter R09b; R11) with the Advanced Camera for Surveys (ACS) Wide Field Channel (WFC) and/or the Wide-Field Camera 3 (WFC3) Ultraviolet \& Visible Channel (UVIS). ACS/WFC has a field of view of $202\arcsec$ on a side sampled at $0\farcs05$~pix$^{-1}$ while WFC3/UVIS has a field of view $162\arcsec$ on a side sampled at $0.04\arcsec$~pix$^{-1}$ \citep{dressel16}. Finally, we obtained new observations of 9 \snh\ using WFC3. We obtained the majority of our optical images with these modern cameras, 113 and 132 unique epochs with ACS and WFC3 (respectively), while WFPC2 contributes a smaller fraction with 67 epochs. Table~\ref{tab:obs} contains information on all the optical observations used in our analysis (both archival or newly obtained) including the proposal ID, camera, date, and exposure time in each filter. Figure~\ref{fig:galaxies} displays a color image of each SN~Ia host galaxy and the field observed with {\it HST}. Additional observations of all targets, obtained using the infrared channel of WFC3, are described and analyzed by R16.

\begin{figure}[htbp]
\begin{center}
\includegraphics[width=0.49\textwidth]{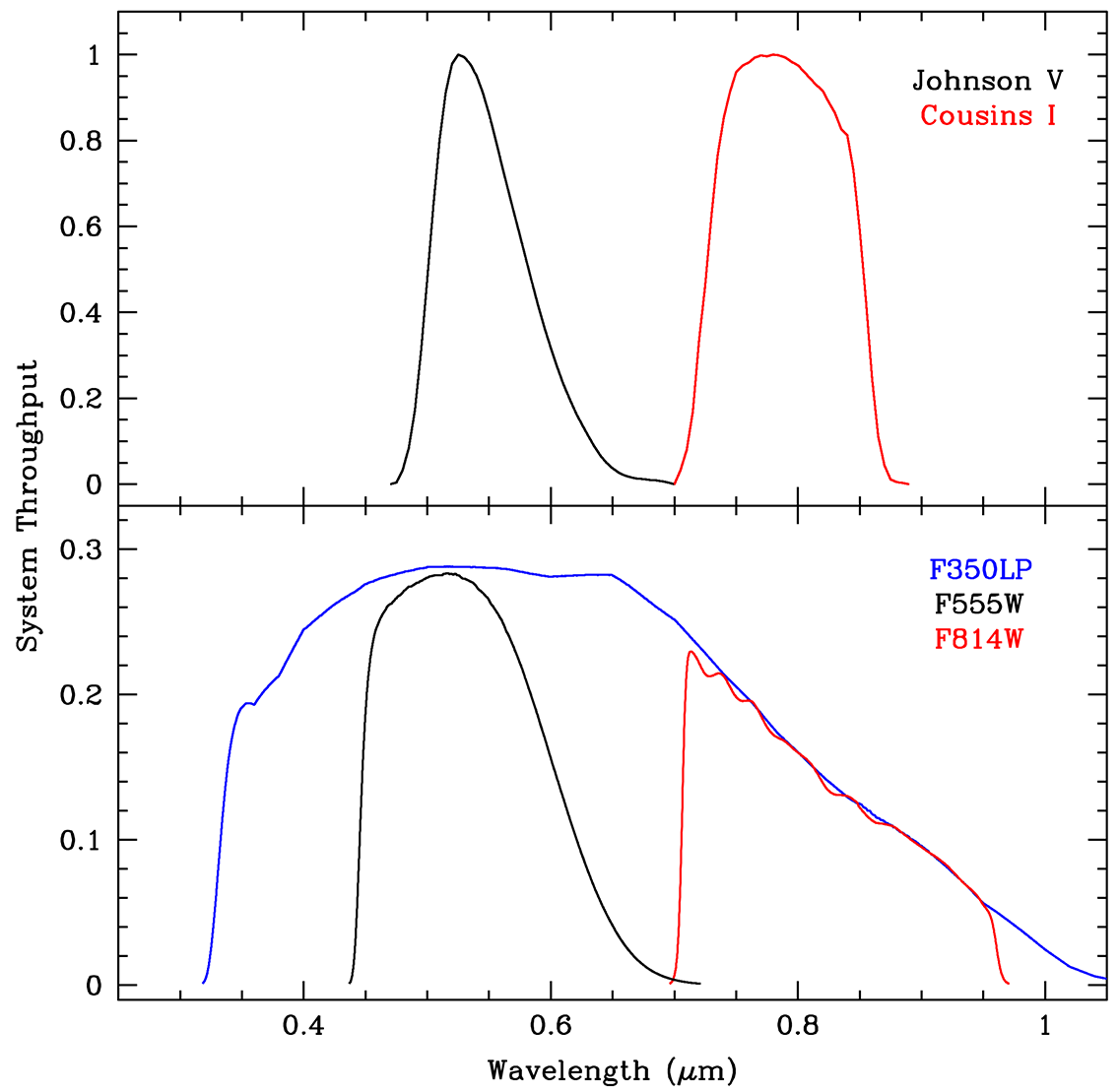}
\caption{Top: filter transmission curves of the traditional $V$ and $I$ filters. Bottom: System throughput (as calculated by {\tt SYNPHOT}) for the {\it HST} filters used in this work.}
\label{fig:filters}
\end{center}
\end{figure}

Given the heavy oversubscription of {\it HST}, it is desirable to minimize the number of orbits required to discover and characterize Cepheids. Therefore, we took advantage of a novel capability on {\it HST} when imaging the new {\snh}: a wide ``white light'' filter (labeled {\W}) available on WFC3/UVIS that enables detection and phasing of these variables $\sim 2.5$ times faster than the traditional {\V} filter. We imaged each target 11--12~times over 60--100~days, depending on roll-angle constraints and the number of orbits required per epoch. In the case of galaxies for which we used the {\W} filter to carry out the Cepheid search, we obtained shallow images on several epochs using the {\V} ($V$) and {\I} ($I$) filters. These images were stacked and used to obtain mean-light $V\!-\!I$ color information (see \S\ref{sec:psfphot}), critical for a consistent selection of Cepheid candidates and for subsequent corrections to the infrared magnitudes. Figure~\ref{fig:filters} shows the wavelength range spanned by these filters.

\begin{deluxetable}{cccccccc}[t]
\tabletypesize{\scriptsize}
\tablecaption{{\it HST} Observations Analyzed in this Work\label{tab:obs}}
\tablewidth{0pc}
\tablehead{\colhead{Gal.} & \colhead{Prop} & \colhead{Cam.} & \colhead{Date} & \multicolumn{3}{c}{Exp.~time (s)} & \colhead{D}\\
\colhead{} & \colhead{ID} & \colhead{} & \colhead{} & \colhead{V} & \colhead{I} & \colhead{W} & \colhead{}}
\startdata
\fgal & 12880 & WFC3 & 2013-06-30 & 480 & 600 & 2288 & A \\
\fgal & 12880 & WFC3 & 2013-07-10 & 480 & 600 & 2288 & A \\
\fgal & 12880 & WFC3 & 2013-07-20 & 480 & 600 & 2288 & A \\
\fgal & 12880 & WFC3 & 2013-07-30 & 480 & 600 & 2288 & A \\
\fgal & 12880 & WFC3 & 2013-08-08 & 480 & 600 & 2288 & A \\
\fgal & 12880 & WFC3 & 2013-08-16 & \nd & \nd & 2288 & A \\
\fgal & 12880 & WFC3 & 2013-08-21 & 480 & 600 & 2288 & A \\
\fgal & 12880 & WFC3 & 2013-08-31 & 480 & 600 & 2288 & A \\
\fgal & 12880 & WFC3 & 2013-09-09 & 480 & 600 & 2288 & A \\
\fgal & 12880 & WFC3 & 2013-09-16 & 480 & 600 & 2288 & A \\
\fgal & 12880 & WFC3 & 2013-09-26 & 480 & 600 & 2288 & A \\
\fgal & 12880 & WFC3 & 2013-10-08 & 480 & 1080 & 2288 & A
\enddata
\tablecomments{V={\V}; I={\I}; W={\W}. D: Images processed using [A]stroDrizzle or [M]ultiDrizzle. This table is available in its entirety in a machine-readable form in the online journal. A portion is shown here for guidance regarding its form and content. } 
\end{deluxetable}
 
The flexibility in scheduling made possible by space-based observations allows for the optimal sampling of Cepheid light curves with a minimal number of observations.  The scheduling strategy for previous {\it HST}-based Cepheid searches has relied upon a logarithmic spacing of observation intervals \citep{madore05}.  This technique, sometimes referred to as a ``power-law time distribution,'' provides good sensitivity to all possible periods within the observational baseline.  However, since the Cepheids we seek have clear upper and lower limits in period, it is possible to select sets of observation time spacings which are superior to those derived from a power-law distribution for our period range.  In effect, the existence of a range of preferred periods {\it does\/} provide a natural scale which we can exploit, unlike the completely scale-free power law. A strategy similar to ours was independently derived by \citet{saunders06,saunders08}.

\begin{figure}[htbp]
\begin{center}
\includegraphics[width=0.49\textwidth]{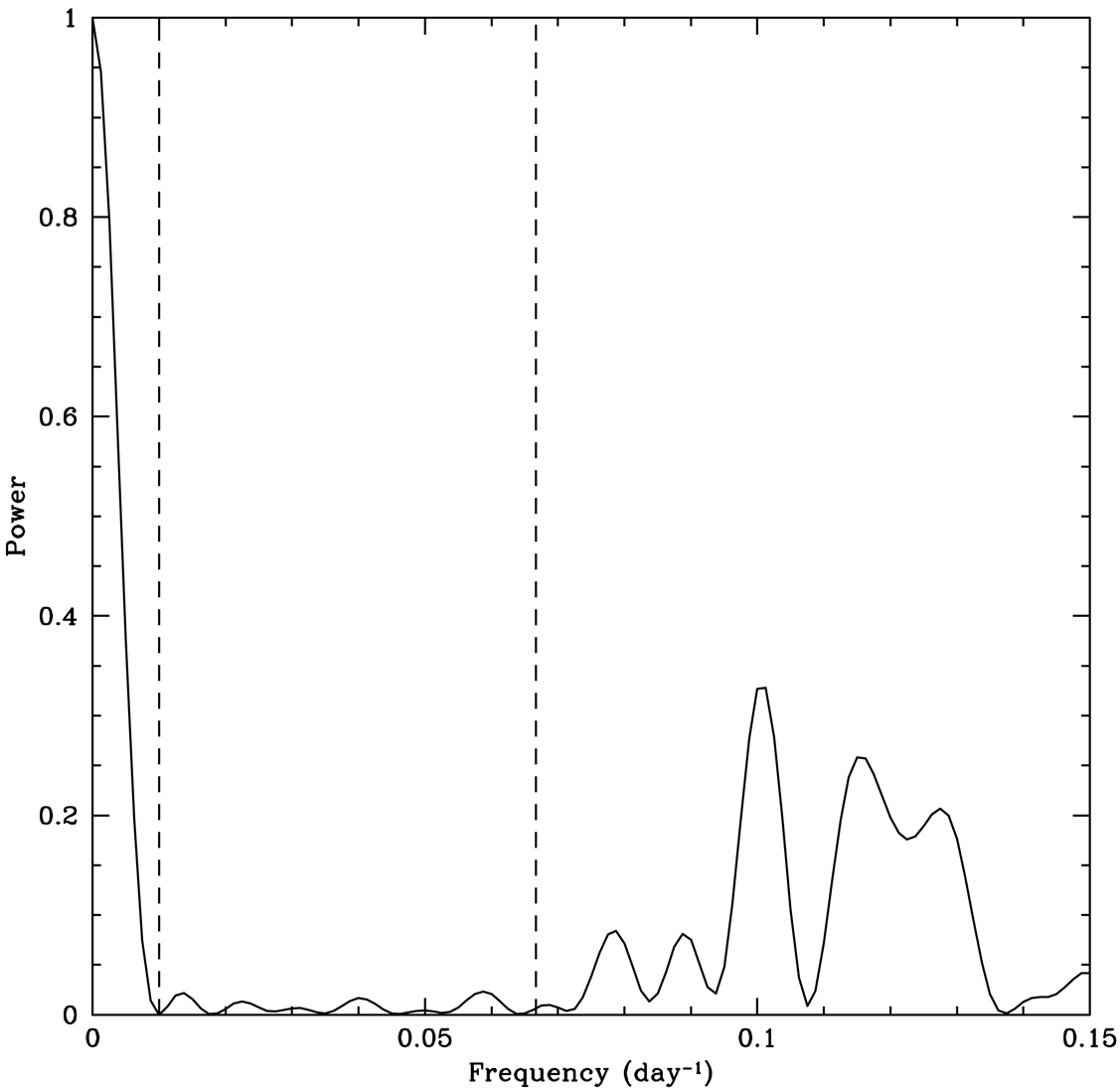}
\caption{The resulting power spectrum for one set of 100~day observations with a cadence chosen to minimize aliasing in the chosen period range. The dotted line representing the power spectrum is lowest between the two dashed lines indicating the 15 and 100~day limits in the frequency.}
\label{fig:alias}
\end{center}
\end{figure}

The assertion that a set of observation time spacings can be selected which will be superior to a power-law distribution is easily demonstrated by examination of the power spectrum derived from a discrete Fourier transform of candidate observation times.  A superior set of observation times will result in lower total power and reduced alias peak size over the range of frequencies which correspond to reasonable Cepheid (inverse) periods.  Both analytic and ``brute force'' prediction algorithms are possible for the production of observation time sequences. In our case, our proposed time interval list was determined by selecting times randomly generated from within an observation window whose upper bound was set by the roll-angle constraint of {\it HST} and whose lower bound was set at 9--10~d, since shorter-period Cepheids are unlikely to be detected at the distances of the galaxies in our sample.  Those spacings which minimized the integral of the power over the frequency interval corresponding to the observation window were retained, as depicted in Figure~\ref{fig:alias}. Alternatively, one could select the observational intervals on the basis of minimizing the maximum amplitude of aliasing features within the observational window, or some weighted combination of area and amplitude of the power spectrum. In any event, the set of observational intervals which result from this procedure is clearly superior to that selected from a power-law distribution. In instances when the originally planned sequence was interrupted by a safing event or a failed acquisition, a set of remaining epochs was produced which, although less optimal than the original sequence, minimized aliasing in the range of expected detectable Cepheid periods. We experienced this only once, when observing {\ugal}. 

We retrieved pipeline-processed images using the Mikulski Archive for Space Telescopes (MAST). In the case of new observations or targets that were never previously analyzed by the SH0ES project, we generated our own ``drizzled'' images for single-epoch and stacked master frames using v1.1.8 of the {\tt AstroDrizzle} package \citep{gonzaga12} with the native WFC3 pixel size and the pixel fraction parameter set to 1. The MAST pipeline provided images that were already corrected for the effects of charge-transfer efficiency (CTE) for the ACS data only. Therefore, we performed CTE corrections for the WFC3 images using v1.0 of the stand-alone {\tt wfc3uv\_ctereverse} program provided by STScI. In the case of galaxies analyzed in previous iterations of our project, we used the existing photometry originally performed on images created with the {\tt Multidrizzle} package \citep{fruchter08} but applied CTE corrections derived from new master frames. The last column of Table~\ref{tab:obs} indicates the procedure followed for each target. In all cases, individual images were registered and aligned to better than $0.1$~pix. 

\begin{figure}[htbp]
\begin{center}
\includegraphics[width=0.49\textwidth]{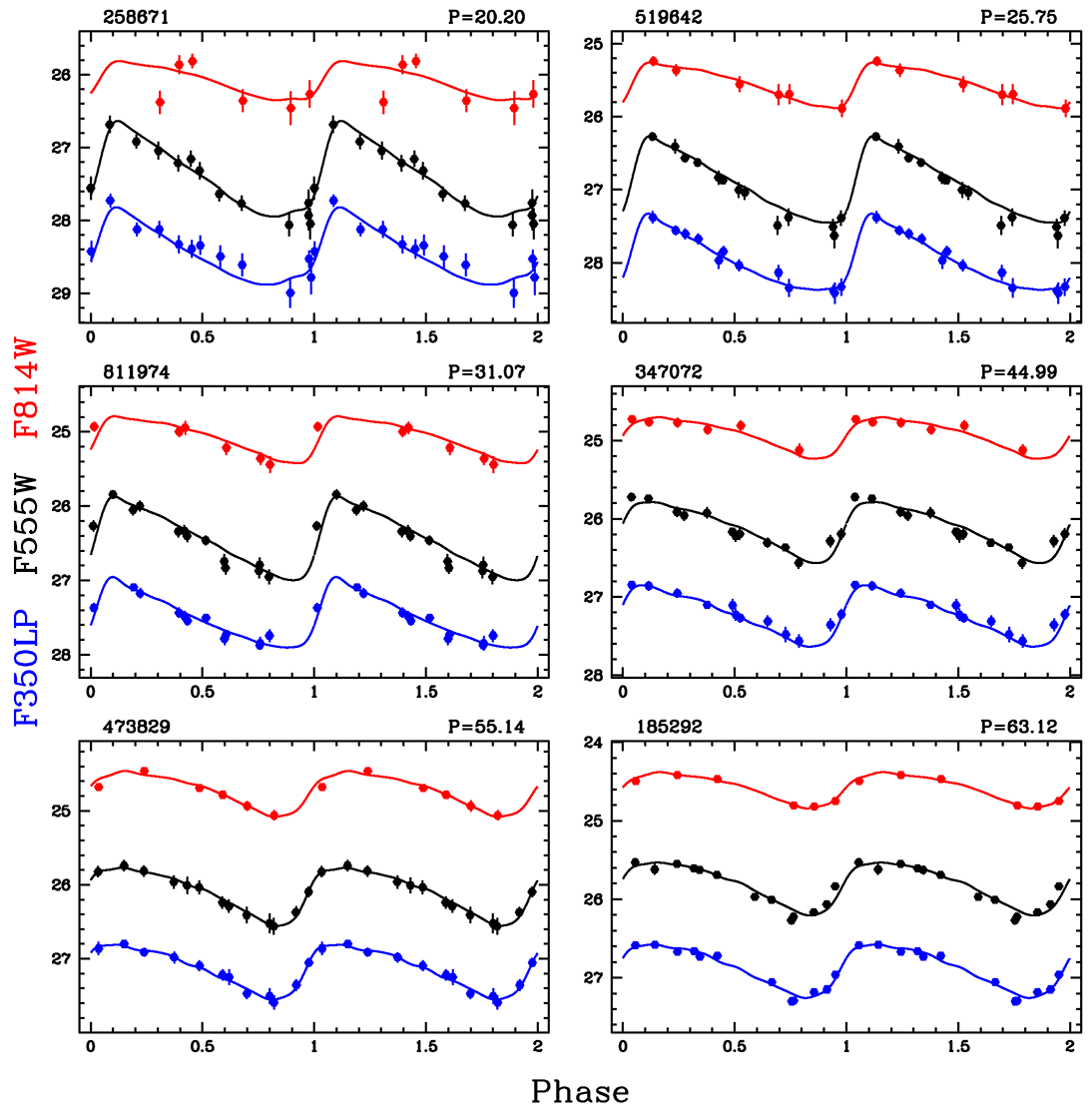}
\caption{Representative light curves of Cepheids in {\tgal} for the {\V} (middle, black), {\I} (top, red), and {\W} (bottom, blue) filters. Two cycles are plotted and offsets of $-0.25$ and +1.25~mag were applied to {\I} and {\W} (respectively) to aid visualization. The best-fit templates from \cite{yoachim09} are plotted using solid lines. The derived periods are given in days in the top-right corner of each panel. All information for these variables is presented in Table~\ref{tab:ceph}.}
\label{fig:n5584lc}
\end{center}
\end{figure}

\begin{figure}
\begin{center}
\includegraphics[width=0.49\textwidth]{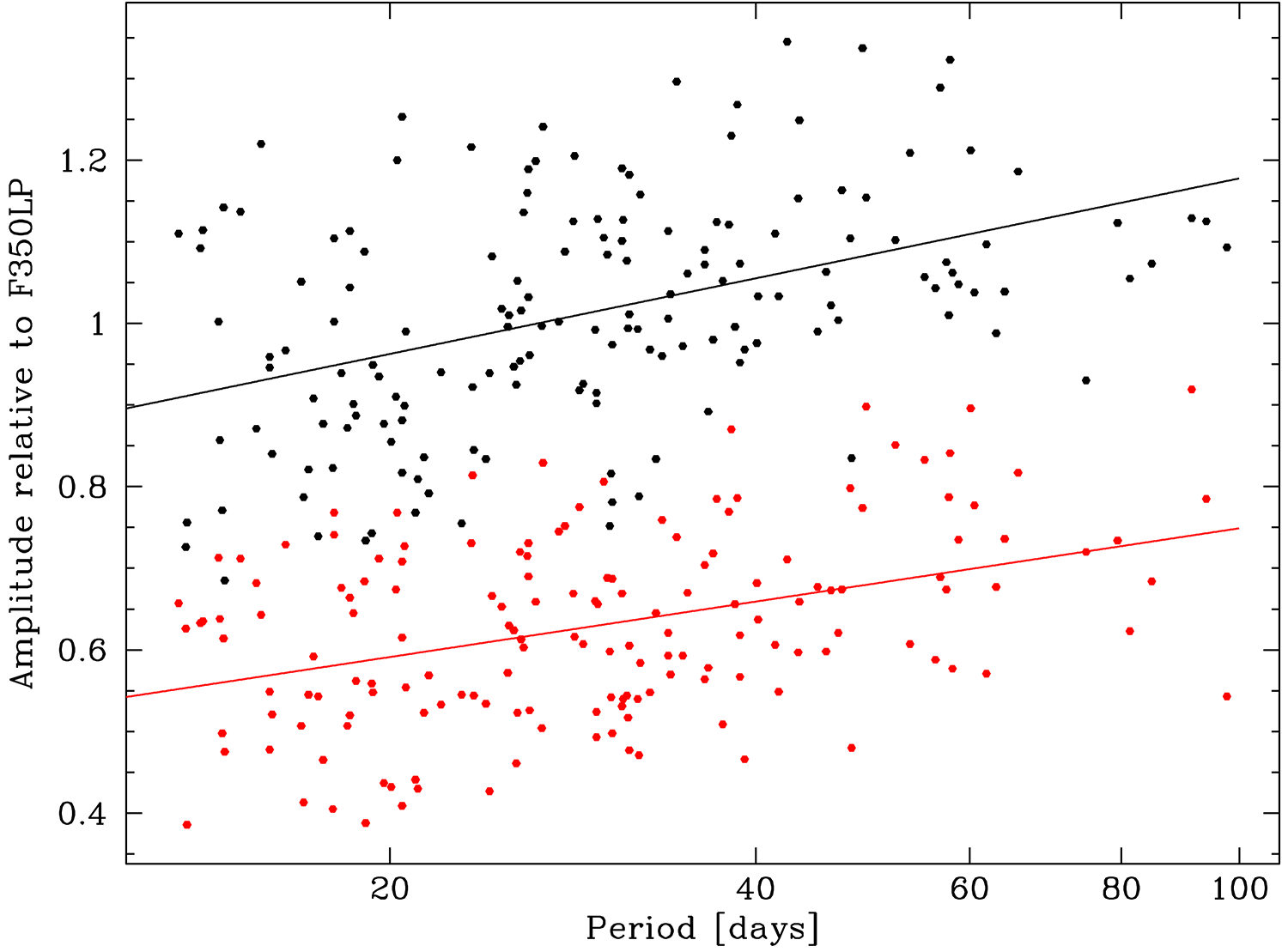}
\caption{Cepheid amplitude ratios in {\tgal}, relative to {\W}, for {\V} (black) and {\I} (red). Linear fits as a function of period are displayed as solid lines. Results are presented in Table~\ref{tab:tgalarat}.}
\label{fig:n5584amprat}
\end{center}
\end{figure}

\section{Photometry}\label{sec:psfphot}

Given the crowded nature of our fields, we performed time-series PSF photometry using {\tt DAOPHOT}, {\tt ALLSTAR}, and {\tt ALLFRAME} \citep{stetson87, stetson94}, a well-established suite of programs extensively used for this type of work \citep{stetson98}. The PSFs for the various cameras and chips were determined using the same software on simulated images created with the {\tt TinyTim} package \citep{krist11}. As previously stated, we only used ACS or WFC3 stacked images to generate the input star lists required by {\tt ALLFRAME} and for the subsequent photometric calibration, including the derivation of aperture corrections. We followed the procedures explained in detail by M06 for the generation of the master input lists and frame-to-frame registration. These were the same procedures carried out by R09b and R11 for the analysis of {\snh} targeted in previous iterations of the SH0ES project (flagged with ``M'' in column 8 of Table~\ref{tab:obs}). We performed completely new time-series photometry for the galaxies that were targeted in this phase of the project: 9 new \snh\ plus 5 hosts that benefited from additional imaging. In the case of galaxies with no new optical imaging relative to our previous publications ({\rgal}, {\agal}, {\lgal}, {\bgal}, \& {\tgal}), we used the existing time-series photometry. However, we emphasize that we obtained new consistent photometric calibrations for {\it all} targets making use of WFC3 {\V} and {\I} images, and we carried out new and consistent procedures for variable-star identification in all \snh\ and Cepheid classification in all galaxies as described below.

As part of the work previously presented by R11, we carried out observations of {\tgal} soon after the installation of WFC3 on {\it HST}. While the search for Cepheids in this galaxy was based on ``traditional'' {\V} and {\I} imaging, a small fraction of many orbits of that campaign was dedicated to {\W} observations. This served as a proof of concept for subsequent ``white-light'' searches and, critically, it allowed us to derive interrelations of Cepheid properties across {\W}, {\V}, and {\I}. Figure~\ref{fig:n5584lc} displays representative light curves of six Cepheids in this galaxy spanning the range of periods covered. It can be seen that the distinct ``saw-tooth'' light-curve shape of Cepheids in {\V} is also present in {\W} and closely matches it in terms of phase and amplitude.

We fit all Cepheid light curves using templates generated by \cite{yoachim09}. We solved for the best-fit amplitude of each {\tgal} variable in each band and derived amplitude ratios (relative to {\W}) as a function of period, shown in Figure~\ref{fig:n5584amprat}. These were later used in the analysis of Cepheids with time-series information only available in {\W}, as explained in the next subsection. We found approximately equal amplitudes in {\V} and {\W} and the expected 0.6:1 ratio of $I/V$ amplitudes \citep{klagyivik09}. We found a weak period dependence of the amplitude ratios, which we modeled using a linear function of $\log P$ since higher-order terms were only significant at the $\sim 1\sigma$ level (see Table~\ref{tab:tgalarat} for details).

\begin{figure}[htbp]
\begin{center}
\includegraphics[width=0.49\textwidth]{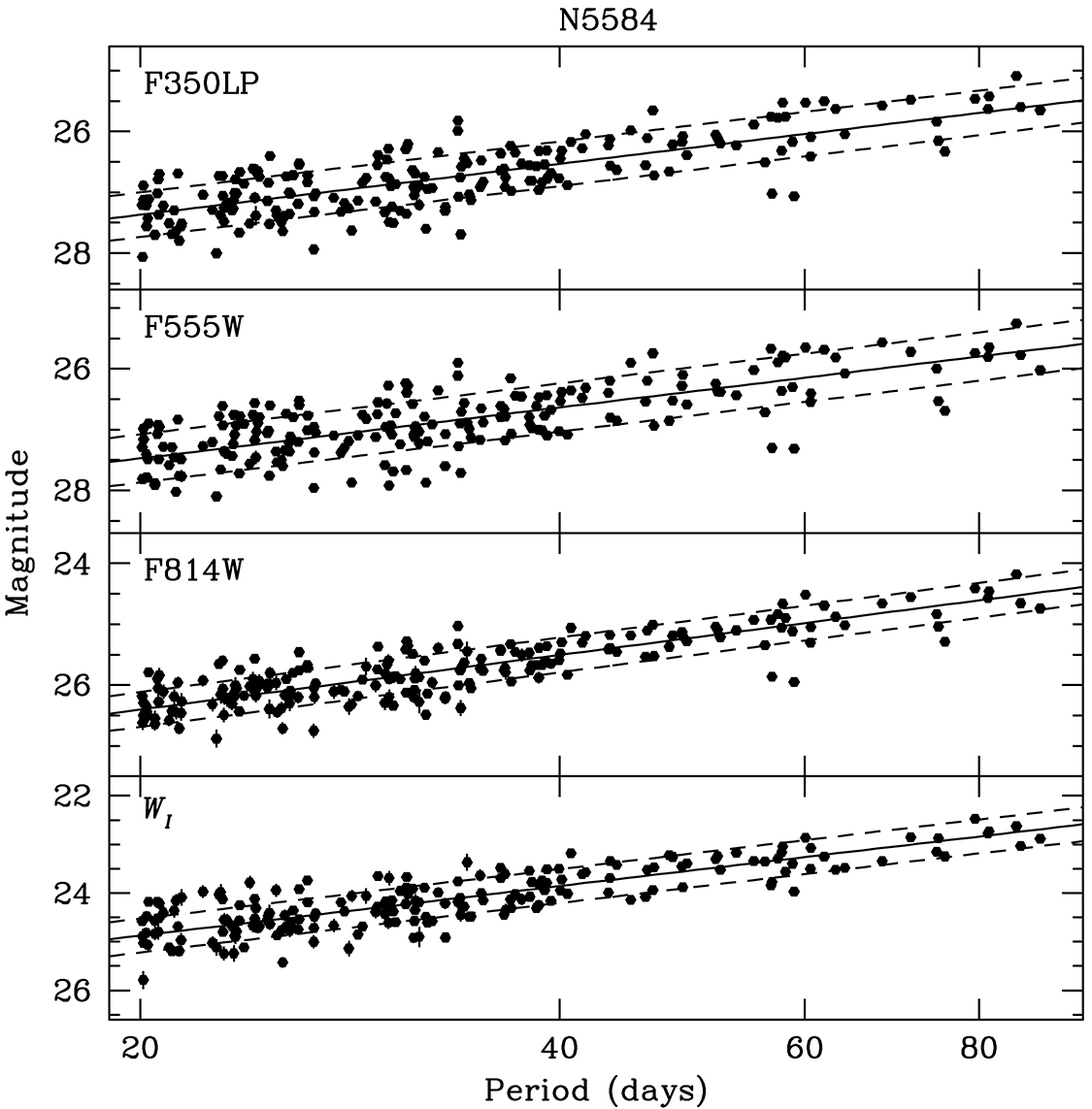}
\caption{Period-luminosity relations in {\W}, {\V}, {\I}, and $W_{I}$ for Cepheids in {\tgal} that comprise the final subsample used by R16. The solid lines represent slopes derived from LMC Cepheids by \cite{udalski99} in $V$ (used for {\W} and {\V}) and $I$ (used for {\I}), and from a global fit by R16 in $W_I$.}
\label{fig:n5584pl}
\end{center}
\end{figure}

Reassuringly, we found that {\W} observations yield a coherent PLR as shown in Figure~\ref{fig:n5584pl}, with a dispersion and a slope similar to the $V$-band PLR. We expect these similarities based on the similar effective wavelengths of Cepheids in these filters (0.53~$\mu$m in {\V} and 0.61~$\mu$m in {\W} for a G0~V star). We emphasize that the {\W} PLR is shown only for illustrative purposes; it is not used for sample selection at any point in our analysis. 

\begin{deluxetable}{lcllc}
\tabletypesize{\tiny}
\tablecaption{Properties of PLRs and Amplitude Ratios in {\tgal} \label{tab:tgalarat}}
\tablewidth{0pc}
\tablehead{\multicolumn{1}{l}{Band} & \multicolumn{1}{c}{$\sigma_{\rm PL}$} & \multicolumn{2}{c}{$A/A_W$} & \multicolumn{1}{c}{$\sigma(A/A_W)$}\\
\colhead{} & \multicolumn{1}{c}{[mag]} & \colhead{$c_0$} & \colhead{$c_1$} & \colhead{}}
\startdata
\V & 0.39 & $1.024\pm0.011$ & $-0.308\pm0.052$ & 0.134 \\
\I & 0.31 & $0.636\pm0.008$ & $-0.226\pm0.041$ & 0.107 \\
\W & 0.37 & \nd  & \nd & \nd 
\enddata
\tablecomments{W={\W}. $A/A_W=c0+c1 (\log P - 1.5)$.}
\end{deluxetable}
 
\begin{figure}
\begin{center}
\includegraphics[width=0.49\textwidth]{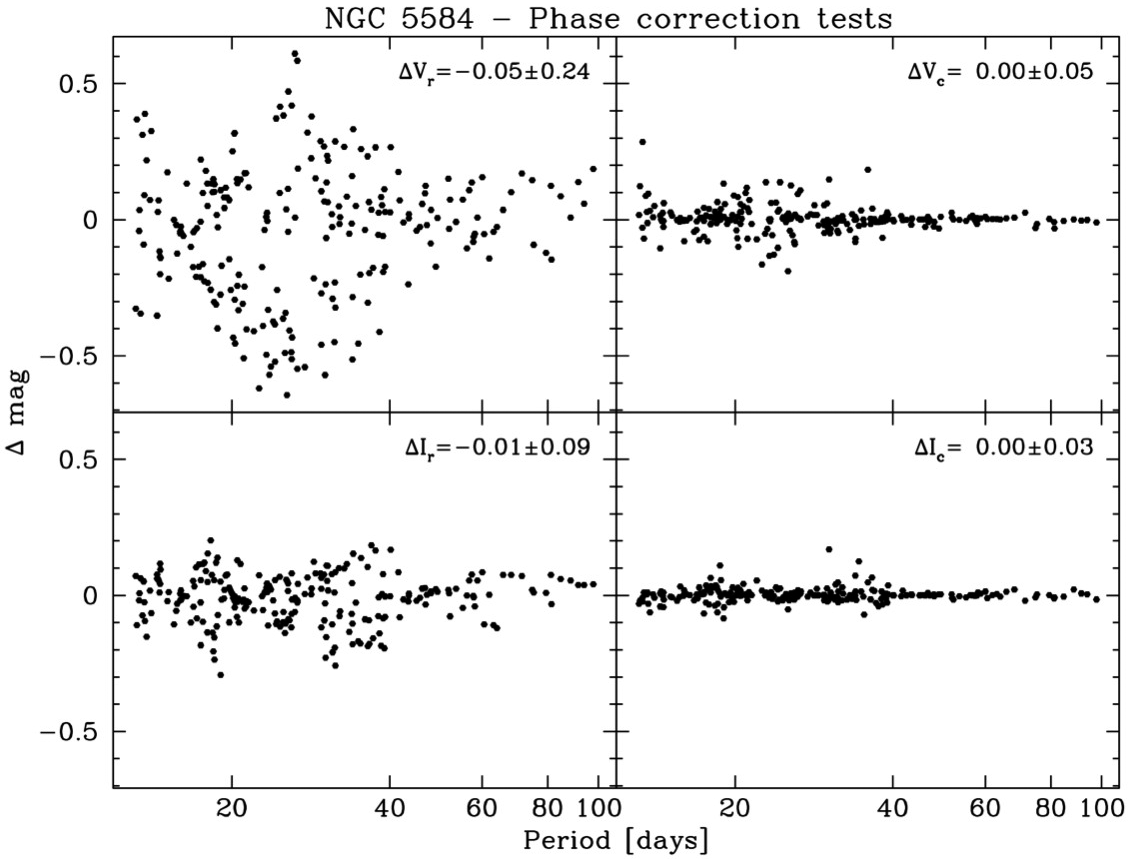}
\caption{Results of the phase correction tests in {\tgal}. Left panels: no correction applied. Right panels: correction applied. The correction does not introduce any statistically significant change in mean magnitudes and helps to drastically reduce the dispersion.}
\label{fig:phctest}
\end{center}
\end{figure}

Having validated the ``white-light'' approach described above, we observed an additional 9 {\snh} using only {\W} for time-series photometry. We corrected the random-phase observations in {\V} and {\I} to mean-light values using the \cite{yoachim09} templates and the relations between amplitude ratio and period derived from the observations of {\tgal}. We tested the procedure by phase-correcting random epochs of {\V} and {\I} photometry of Cepheids in {\tgal} using only the information from the {\W} light curves, and comparing the result with the mean magnitude derived from the template fits to the full {\V} and {\I} light curves. The results are shown in Figure~\ref{fig:phctest}; the left panels demonstrate the large dispersion that results from random-phase observations, while the right panels illustrate the significant improvement realized with our correction procedure, resulting in uncertainties of $\sigma_{V}\!=\!0.05$ and $\sigma_{I}\!=\!0.03$. We emphasize that this procedure relies only on the assumption that the amplitude ratios found in {\tgal} are applicable to the {\snh} observed primarily in {\W}, and not on whether Cepheid amplitudes at a given period are independent of metallicity \citep[see][]{szabados12,majaess13}. In any event, the Cepheids corrected by this procedure have abundances very similar to those used to derive the correction (see \S\ref{sec:chem} and Table~\ref{tab:ceph}; $\langle[{\rm O/H}]\rangle\!=\!8.84\pm0.14$~dex for {\tgal} and $\langle[{\rm O/H}]\rangle\!=\!8.88\pm0.22$~dex for the 9 ``white-light'' SN~Ia hosts).

We further corrected the mean {\V} and {\I} PSF magnitudes of all stars to the standard apertures of WFC3 ($0\farcs4$) and ACS ($0\farcs5$). We derived growth curves for each detector and filter following the standard approach of selecting bright, isolated stars across all frames, removing all other objects from the images through PSF subtraction, and carrying out aperture photometry at a variety of radii between $0\farcs15$ and the values listed above. We found the growth curves for all individual frames of a given detector and filter to be quite consistent with each other, and therefore averaged them to improve the robustness of this correction. The only exception was M101, where the larger number of stars enabled a separate determination of the growth curves which differed slightly (0.01~mag in {\V} and 0.03~mag in {\I}).

All galaxies were observed with WFC3 {\V} and {\I} to provide a consistent set of photometric zeropoints. For a few cases, the ACS data were significantly deeper, so we combined ACS and WFC3 magnitudes using transformations derived with {\tt SYNPHOT} \citep{laidler08}, which uses the well-characterized throughput information of the ACS and WFC3 filters. We computed synthetic magnitudes using six stellar-atmosphere models from \citet{castelli04} that were closest to our Cepheids: solar metallicity, $2.1\!<\!\log g\!<\!2.6$, and $4100\!<\!T_{\rm eff}~({\rm K})\!<\!5150$ (equivalent to $0.85\!<\!V\!-\!I\!<\!1.4$~mag). We determined zeropoint offsets ($\Delta$mag $[$ACS$-$WFC3$]$) of $-0.052$ and $+0.829$ for {\V} and {\I}, respectively; the large value in {\I} reflects the lower quantum efficiency of the WFC3 detector at these wavelengths. The offsets have uncertainties of $0.001$~mag, estimated from the scatter of the synthetic magnitudes about the mean value. Finally, we applied the UVIS 2.0 WFC3 Vegamag zeropoints of $25.741$ and $24.603$ for {\V} and {\I}, respectively \citep{bowers16}, and the crowding corrections described in \S\ref{sec:crowd}, to obtain fully calibrated magnitudes. 

\section{Search for Cepheid Variables} \label{sec:idceph}

The time-series photometry of all target galaxies presented in this paper was subject to a new search for Cepheids with improved template-based period determinations and universal selection criteria. The motivation behind this effort was to obtain a homogeneous sample that minimized selection bias. 

\subsection{Identification of Variable Objects }\label{sec:varobjs}

We identified variable objects in our time-series photometry using the Welch-Stetson variability index $L$ \citep{stetson96}, determined by the {\tt TRIAL} program kindly provided by P.~Stetson. The calculation requires the derivation of epoch-to-epoch zeropoint offsets that are based on the error-weighted mean magnitudes of bright, isolated stars (hereafter, ``local standards''). We selected these objects through visual inspection and iteratively discarded those that exhibited variability or had unusually large photometric errors as reported by {\tt ALLFRAME}, and we list the position and magnitudes of these local standards in Table~\ref{tab:secstds}. The error-weighted mean magnitudes for each filter were based on the WFC3 or in a few cases ACS time-series photometry for a given galaxy, occasionally excluding epochs with large zeropoint offsets that arose from defocusing or imperfect guide-star lock.

\begin{deluxetable}{llllrrrr}
\tablecaption{Local standard stars \label{tab:secstds}}
\tablewidth{0pc}
\tablehead{\colhead{Gal.} & \colhead{ID} & \colhead{$\alpha$} & \colhead{$\delta$} & \multicolumn{4}{c}{Magnitude} \\ \colhead{} & \colhead{} & \multicolumn{2}{c}{(J2000, deg)} & \colhead{V} & \colhead{$\sigma$} & \colhead{I} &\colhead{$\sigma$}}
\startdata
\mgal &  190735 & 210.89053 &  54.37253 & 21.333 &   9 & 20.966 &   9 \\
\mgal &  329066 & 210.85711 &  54.37351 & 21.459 &  10 & 21.194 &   7 \\
\mgal &  258240 & 210.86774 &  54.36504 & 21.630 &  11 & 21.655 &   6 \\
\mgal &  240467 & 210.85522 &  54.34199 & 21.873 &  17 & 20.759 &  12
\enddata
\tablecomments{V={\V}; I={\I}. $\sigma$ expressed in mmag. This table is available in its entirety in a machine-readable form in the online journal. A portion is shown here for guidance regarding its form and content. }
\end{deluxetable}

\begin{figure}
\begin{center}
\includegraphics[width=0.49\textwidth]{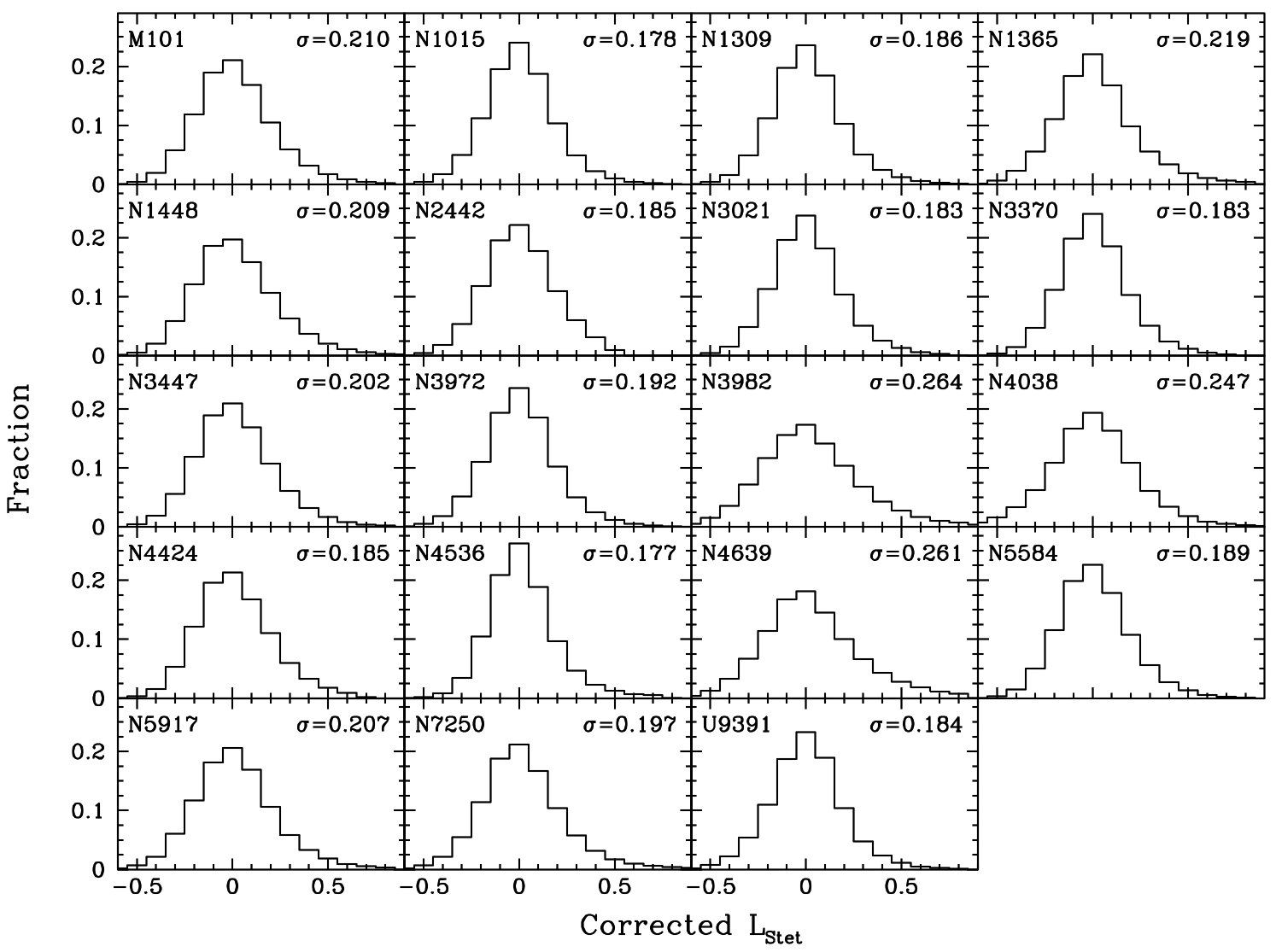}
\caption{Distribution of {\it L} values for all stars in each SN~Ia host, after correcting the photometric uncertainties originally reported by ALLFRAME (see text for details). The dispersion for each galaxy is given at the top right of each panel.}
\label{fig:lv}
\end{center}
\end{figure}

 It is well known \citep[see \S4.3 of][]{kaluzny98} that the photometric errors reported by {\tt DAOPHOT} and related programs require a magnitude-dependent rescaling to yield consistent variability indices. We applied this correction and then flagged as variables all objects with $L\geq 0.75$. Given the relatively small samples in {\wgal} and {\ogal}, we lowered the $L$ threshold to 0.60 and 0.65, respectively, to examine additional light curves. However, we found only 5 and 1 additional candidates, respectively, that passed the visual inspection. Figure~\ref{fig:lv} shows the distribution of $L$ versus magnitude for all fields. 

\begin{figure}[htbp]
\begin{center}
\includegraphics[width=0.49\textwidth]{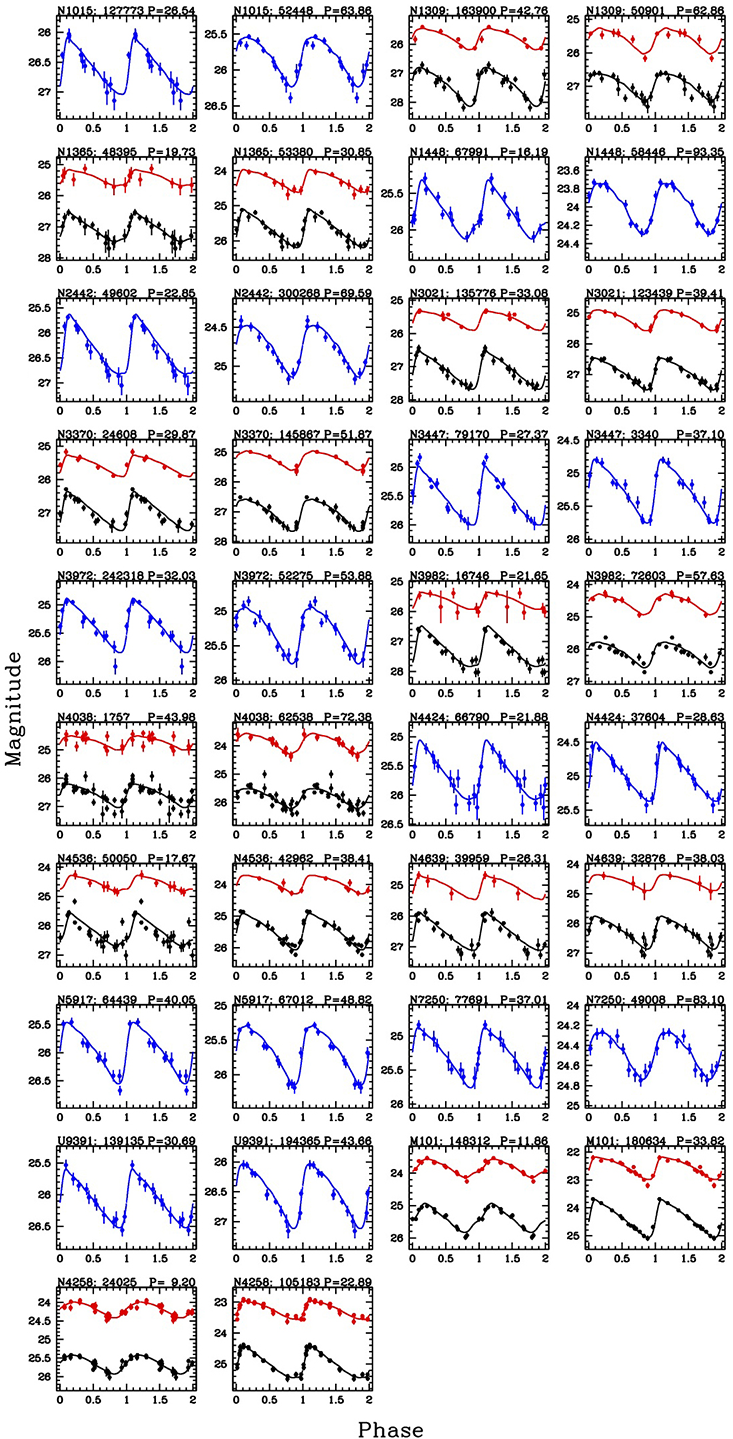}
\caption{Representative Cepheid light curves and best-fit templates for each of the galaxies analyzed in this work. {\W} is plotted in blue, {\V} in black, and {\I} in red. The latter was offset by $-1.5$~mag for clarity.}
\label{fig:megalc}
\end{center}
\end{figure}

In the case of \ngal, we benefited from extensive additional resources from previous Cepheid searches and did not carry out a new search for variables. We used the Cepheid candidates from M06 as well as the ground-based samples from \cite{fausnaugh15} and \cite{hoffmann15}, which have considerably more extended baselines than the typical {\it HST}-based search. Hence, the latter two studies preferentially detected long-period Cepheids and aided our efforts to match the characteristics of the Cepheid population in this galaxy to those in the {\snh}. We identified the \cite{hoffmann15} Cepheids in a single-epoch galaxy-wide mosaic obtained as part of this project using {\it HST} ACS/WFC {\V} and {\I} images and phase-corrected their magnitudes to mean light. \cite{fausnaugh15} had already calculated the {\it HST} {\V} and {\I} mean magnitudes as one of the intermediate steps in their analysis, and kindly provided the measurements to us.

\begin{figure}[htbp]
\begin{center}
\includegraphics[width=0.49\textwidth]{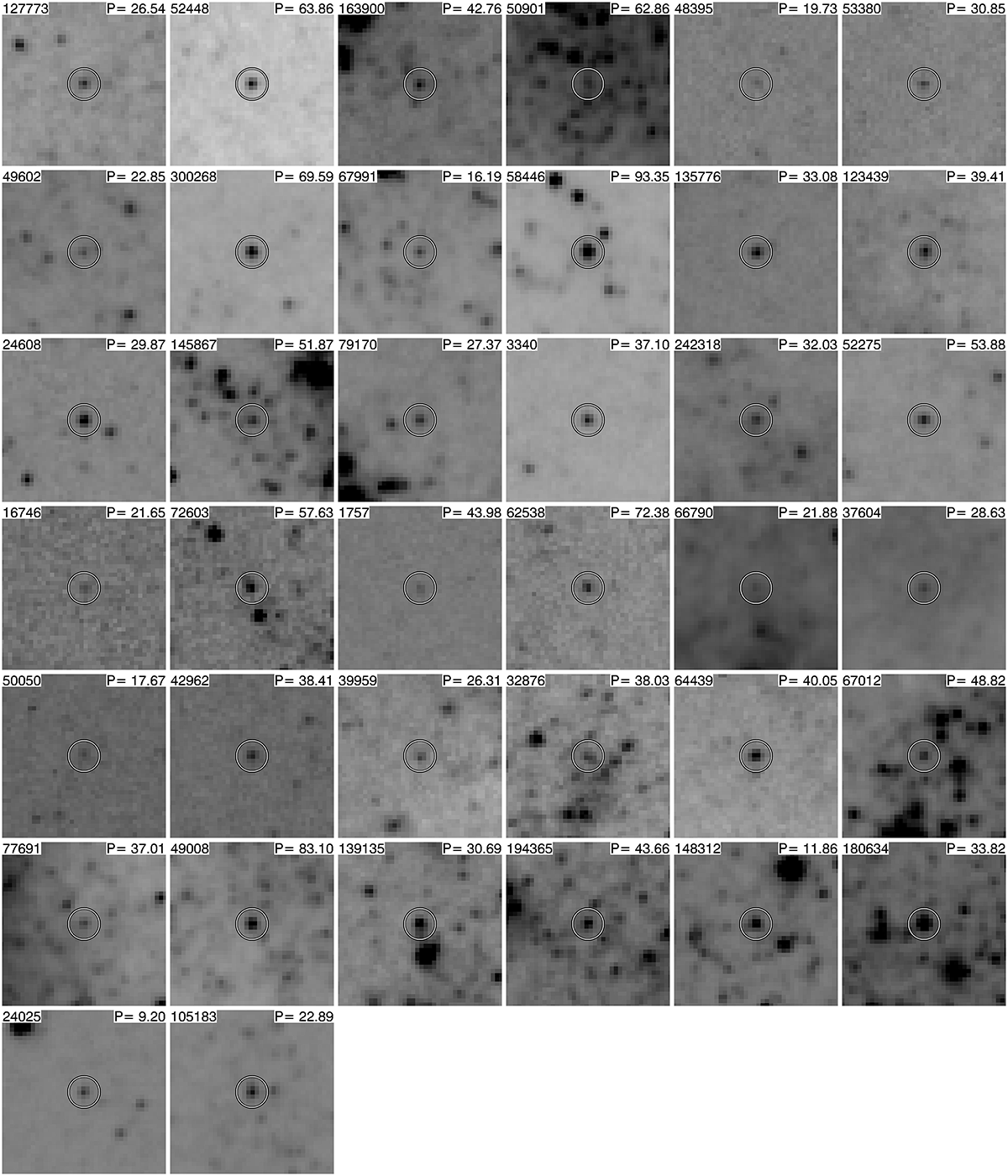}
\caption{Thumbnails of representative Cepheids in each galaxy, matching the light curves plotted in Figure~\ref{fig:megalc}. The deepest master frame (in {\W} or {\V}, as appropriate) was used. Each panel is $6\arcsec$ on a side.}
\label{fig:megalcfc}
\end{center}
\end{figure}

\vfill

\subsection{Cepheid Selection}\label{sec:cuts}

We fit all variable objects with Cepheid light-curve templates \citep{yoachim09} using 150 trial values that were equally spaced in $\log P$ in the range $P=10$--100~d (except for {\mgal} and {\ngal}, where the number of trials was increased and the range lowered to $P_{\rm min}=4$~d owing to their considerably closer distance). We carried out simultaneous light-curve fits in all bands whenever a target had time-series photometry in multiple filters, discarding objects that were undetected at one or more wavelengths. For each initial trial value, the light-curve fitter found the best overall period and phase offset and solved for the light-curve amplitude in each band. We stored the outcome of each trial, including the value of $\chi^2$ based on the fit to the Cepheid template in the ``primary'' band ({\W} or {\V}). 

\begin{deluxetable*}{l|rrrrrrr|rrrrr}
\tablecaption{Universal Criteria Applied for Selection of Cepheid Variables \label{tab:cut}}
\tablewidth{0pc}
\tablehead{\colhead{Galaxy} & \multicolumn{7}{c}{This work} & \multicolumn{5}{c}{R16}\\
\colhead{} & \colhead{Initial} & \colhead{B} & \colhead{R} & \colhead{$\sigma_{VI}$}& \colhead{$W_I$} & \colhead{$P_{\rm min}$} & \colhead{Passed} & \colhead{$W_I$} & \colhead{No H} & \colhead{$W_H$} & \colhead{Final $I$} & \colhead{Final $H$}}
\startdata
\mgal$^{*}$ &  791      &  -75 & -50 &  \nd &  \nd &  -15 &  651 & -136 & -224 & -181 & 515 & 246 \\ 
\fgal &   72           &  -13 &  -7 &  \nd &  \nd &  -21 &   31 &   -6 &  \nd &  -17 &  25 &  14 \\ 
\xgal &  186           &   -9 &  -9 &  \nd &   -7 &  -88 &   73 &  -13 &   -1 &  -28 &  60 &  44 \\ 
\kgal &  101           &   -3 &  -4 &  \nd &   -1 &  \nd &   93 &  -18 &  -15 &  -46 &  75 &  32 \\ 
\ggal &  197           &  -41 & -27 &  \nd &   -3 &  -25 &  101 &  -18 &   -8 &  -39 &  83 &  54 \\ 
\egal &  681           &  -41 & -87 &  -25 &  -40 &  -33 &  455 &  -25 &  -87 & -227 & 430 & 141 \\ 
\ygal &   52           &   -3 &  -1 &  \nd &  \nd &   -5 &   43 &  -27 &   -2 &  -23 &  16 &  18 \\ 
\zgal &  151           &   -2 &  -3 &  \nd &   -6 &  -46 &   94 &  -34 &   -4 &  -27 &  60 &  63 \\ 
\pgal &  239           &  -37 & -18 &  \nd &   -2 &  -36 &  146 &  -34 &  -21 &  -45 & 112 &  80 \\ 
\qgal &  187           &  -41 & -36 &   -3 &  -20 &   -8 &   79 &   -2 &   -2 &  -35 &  77 &  42 \\ 
\rgal &   51           &  -10 &  -1 &  \nd &   -2 &  -14 &   24 &  \nd &  \nd &   -8 &  24 &  16 \\ 
\agal &   41           &   -2 & \nd &  \nd &   -3 &   -1 &   35 &   -5 &   -3 &  -19 &  30 &  13 \\ 
\ngal$^{\dagger}$ & 549  &  -50 & -29 &  \nd &  -69 &  -66 &  335 &  -102 &  -35 & -161 & 233 & 139 \\ 
\wgal &   59           &  -14 & -22 &   -2 &   -7 &   -2 &   12 &   -6 &   -1 &   -8 &   6 &   3 \\ 
\lgal &   57           &   -3 &  -3 &  \nd &   -1 &  \nd &   50 &   -6 &   -2 &  -15 &  44 &  33 \\ 
\bgal &   59           &   -7 &  -5 &  \nd &   -3 &   -5 &   39 &  -22 &   -2 &  -12 &  17 &  25 \\ 
\tgal &  352           &   -2 &  -2 &  \nd &   -4 & -132 &  212 &  -18 &  -13 & -116 & 194 &  83 \\ 
\ogal &   59           &  -11 & -17 &   -3 &   -2 &   -5 &   21 &   -9 &  \nd &   -8 &  12 &  13 \\ 
\dgal &   74           &  -14 &  -9 &   -2 &   -3 &   -2 &   44 &  -20 &  \nd &  -22 &  24 &  22 \\ 
\ugal &   91           &  -14 & -18 &   -2 &  \nd &  -21 &   36 &  -11 &  \nd &   -8 &  25 &  28 \\ 
\hline
Total & 4049  & -392 & -348 & -37 & -173 & -525  & 2574  & -512  & -420  & -1045 & 2062  &  1109 
\enddata
\tablecomments{Criteria applied for removal in this work: $B=V\!-\!I<0.5$~mag; $R=V\!-\!I>1.5$~mag (2.0~mag for \egal, \dgal); $\sigma_{VI}=\sigma(V\!-\!I)$ exceeded maximum value (0.4~mag for $P<25$, 0.3~mag otherwise); $W_I=\ >3\sigma$~outlier in $W_I$ PLR; $P_{\rm min}=$~below minimum period with complete filling of instability strip. *: 33 likely Pop~II pulsators already removed. $^{\dagger}$: We identify some Cepheids from the literature without the necessary information to uniformly apply these criteria, but do so where available. R16 values applicable to $W_I$ ``NML'' variant and $W_H$ ``preferred'' three-anchor solutions: $W_I=\ >2.7\sigma$~outlier in global fit to $W_I$ PLR; {\it No H}: where optically identified Cepheids are not observed in the {\irh} band; $W_H=\ >2.7\sigma$~outlier in global fit to $W_H$ PLR.}
\end{deluxetable*}

\begin{deluxetable*}{lrrrrrrrrrrrrrr}
\tabletypesize{\tiny}
\tablecaption{Cepheid Properties \label{tab:ceph}}
\tablewidth{0pc}
\tablehead{\colhead{Galaxy} & \colhead{$\alpha$} & \colhead{$\delta$} & \colhead{ID} & \colhead{Period} & \multicolumn{4}{c}{Mean Magnitude} & \multicolumn{3}{c}{Amplitude}              & \colhead{Z}     & \colhead{Flag} & \colhead{Lit} \\
           \colhead{}       & \multicolumn{2}{c}{(J2000)}       & \colhead{}   & \colhead{}    & \colhead{\V} & \colhead{$\sigma$} & \colhead{\I}        & \colhead{$\sigma$} & \colhead{\V} & \colhead{\I} & \colhead{\W} & \colhead{} & \colhead{}     & \colhead{} \\ \colhead{}       & \multicolumn{2}{c}{[deg]}       & \colhead{}   & \colhead{[d]}    & \colhead{[mag]} & \colhead{[mmag]} & \colhead{[mag]}        & \colhead{[mmag]} & \multicolumn{3}{c}{[mag]} & \colhead{[dex]} & \colhead{}     & \colhead{} }
\startdata
\fgal &  39.53598 &  -1.33722 &    54744 &  26.096 & 28.012 & 122 & 27.010 & 132 & \nd & \nd & 1.076 & 8.400 &   O &      \\
\fgal &  39.54914 &  -1.30783 &    61919 &  26.171 & 27.954 & 153 & 26.458 & 114 & \nd & \nd & 0.688 & 8.865 &   O &      \\
\fgal &  39.53734 &  -1.32133 &    29667 &  26.301 & 27.463 & 110 & 26.551 &  85 & \nd & \nd & 1.088 & 9.141 &   O &      \\
\fgal &  39.55384 &  -1.32662 &   127773 &  26.539 & 27.950 & 147 & 26.768 & 120 & \nd & \nd & 0.958 & 9.070 &   O &      \\
\fgal &  39.54168 &  -1.32376 &    56181 &  26.835 & 27.866 & 133 & 26.699 &  99 & \nd & \nd & 0.808 & 9.138 &   O &      
\enddata
\tablecomments{$Z=12+\log[{\rm O/H}]$. Flag: indicates Cepheids used in R16 for near-infrared analysis only (H), optical analysis only (O), both (OH), or the 5 variables in {\mgal} used by R16 in the near-infrared analysis with $V\!-\!I$ values outside our limits (X). Lit: indicates a match to a previously published variable; S99 = \cite{silbermann99}, M06 = \cite{macri06}, R11 = \cite{riess11}, S11 = \cite{shappee11}, F15 = \cite{fausnaugh15}, H15 = \cite{hoffmann15}. This table is available in its entirety in a machine-readable form in the online journal. A portion is shown here for guidance regarding its form and content. }
\end{deluxetable*}

\begin{figure*}[htbp]
\begin{center}
\includegraphics[width=\textwidth]{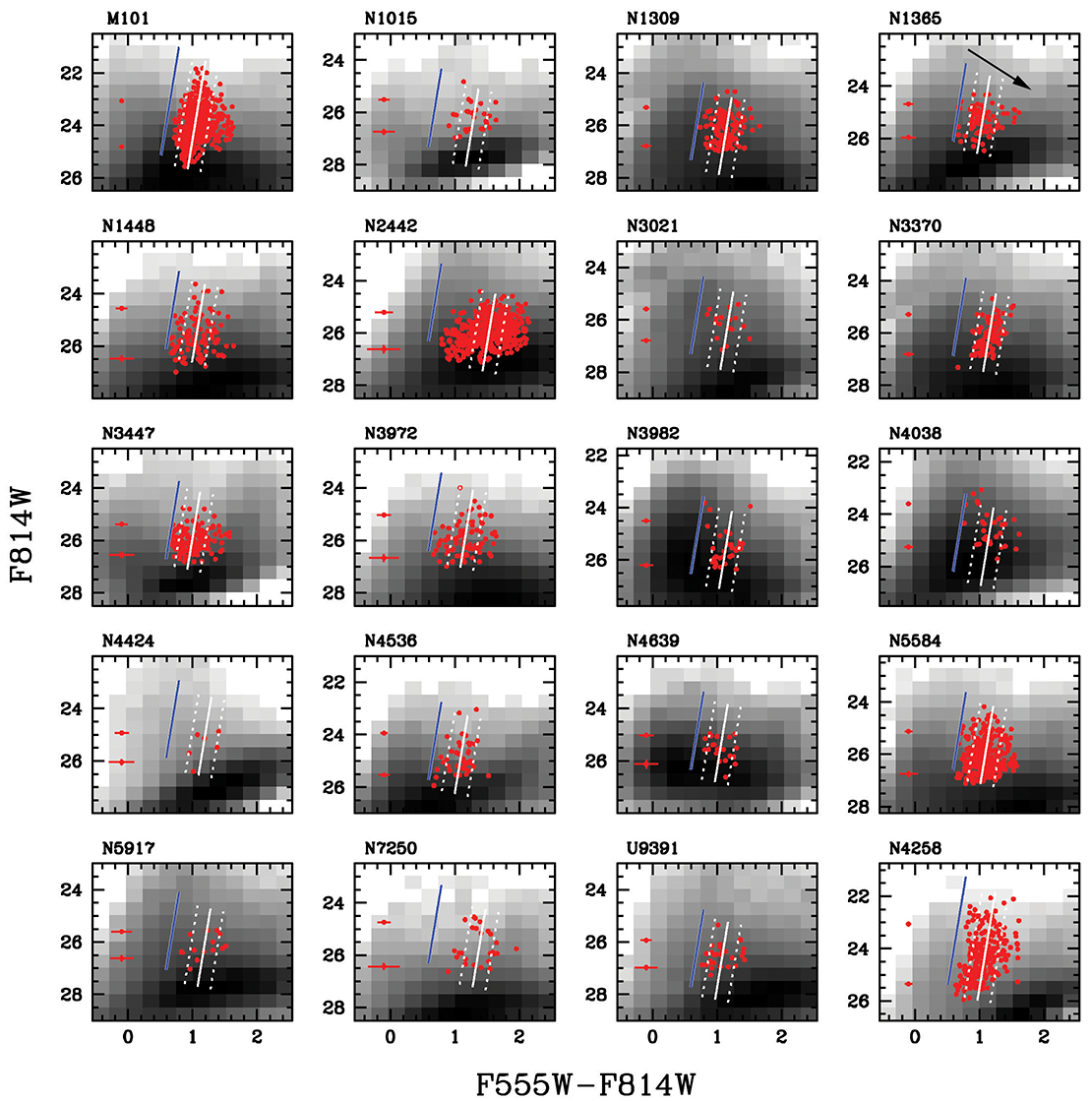}
\caption{Color-magnitude diagrams of the Cepheids (filled red symbols) that passed all the selection criteria (except for minimum period cut, which was not applied to show the complete sample). Representative error bars are shown for the top and bottom quintile of each sample at {\V}$-${\I}$=-0.1$. Ensemble photometry is displayed as Hess diagrams (darker grays convey increased density). The solid blue line shows the $2\sigma$~``blue'' edge of the instability strip in the absence of extinction (derived from LMC Cepheids), shifted to the distance of each galaxy as reported in Table~5 of R16. The solid white line shows the center of the instability strip and the dashed white lines show the $2\sigma$ ``blue'' and ``red'' edges of the instability strip as derived from LMC Cepheids, shifted by the mean color excess of each sample. The arrow plotted in the panel for NGC$\,$1365 shows the effect of $E(V\!-\!I)=1$~mag.}
\label{fig:megacmd}
\end{center}
\end{figure*}
 
\begin{figure}[htbp]
\begin{center}
\includegraphics[width=0.49\textwidth]{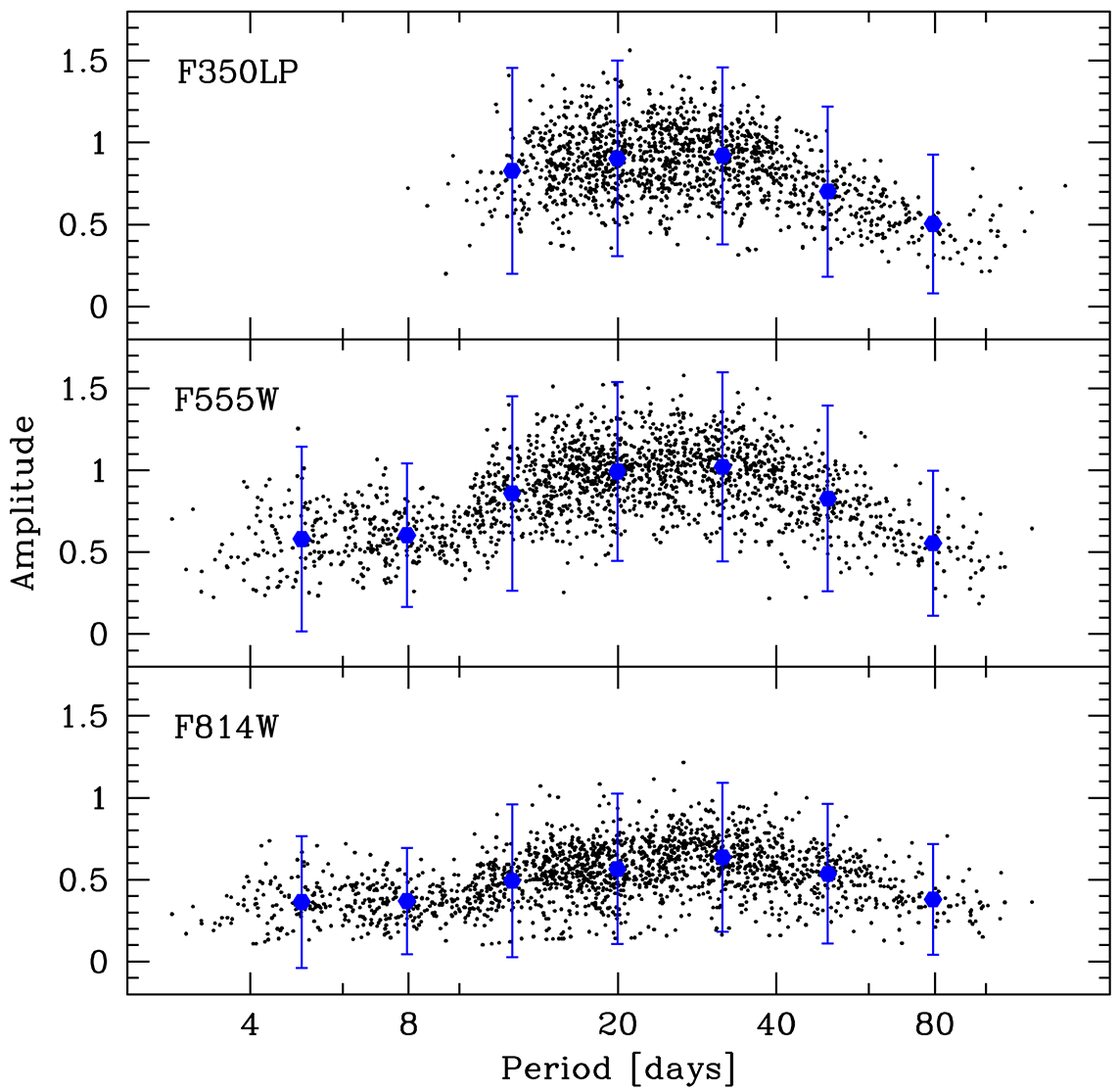}
\caption{Best-fit light-curve amplitudes as a function of period in {\W}, {\V}, and {\I} for the Cepheids that passed all the selection criteria (except for minimum period cut, which was not applied to show the complete sample). Filled blue symbols denote the mean values every 0.2~dex in $\log P$ and the vertical lines denote the $3\sigma$ range.}
\label{fig:megamp}
\end{center}
\end{figure}
 
We visually inspected the six best solutions (in terms of lowest $\chi^2$) for every single variable. The vast majority of the objects in each galaxy (91--98\%) were very poorly fit by the Cepheid template at any trial period, and were removed from further consideration. We cannot reliably reject candidates based on $\chi^2$ because factors other than goodness of fit influence those values. For instance, often Cepheid photometry achieves high S/N, particularly for long-period objects, and thus real substructure in the light curves will artificially inflate $\chi^2$ despite the template fitting the data better than for other variables with lower S/N. Rarely, for objects that passed our visual inspection, we found two solutions with very similar periods and values of $\chi^2$, in which the one with a slightly higher value of the statistic yielded a better fit to data in a secondary band (typically {\I}) or a better overall phase offset. In those cases we selected the better fit despite the small statistical difference. Figure~\ref{fig:megalc} shows sample light curves of Cepheids in each galaxy to display the template fits and the range of periods covered in this analysis. While the templates may not provide a perfect fit to the light curves, the residuals show no statistically significant bias in the derived mean magnitudes. In particular, the color term of the $W_H$ magnitude primarily used by R16 to determine {\ho}, which is based on our {\V} and {\I} measurements, shows a completely negligible offset of $0.002\pm0.002$~mag. We also provide finder charts of the representative Cepheids in Figure~\ref{fig:megalcfc}, from the stacked master images in the primary band of their respective hosts.

We then applied additional selection criteria to remove candidates that, while variable and periodic in nature, failed to meet the expected properties of isolated Cepheids with low to moderate reddening.  We computed the mean {\V} and {\I} magnitudes of each object, either by integrating the best-fit template light curves when we had time-series photometry in those bands or by correcting the random-phase observations to mean light based on the observed {\W} amplitudes and the relations derived in {\tgal}. We further restricted the sample based on the uncertainty in the phase-corrected color, setting limits of $\sigma_{VI}<0.4$~mag (0.3~mag) for variables with $P<25$~d ($>25$~d). We allowed for higher dispersion in the shorter-period (hence, fainter) objects to account for their larger photometric uncertainties. We also applied this cut to variables with time-series photometry in these bands, but no objects were rejected thanks to the statistically more robust determination of mean colors.

\begin{figure*}[htbp]
\begin{center}
\includegraphics[width=\textwidth]{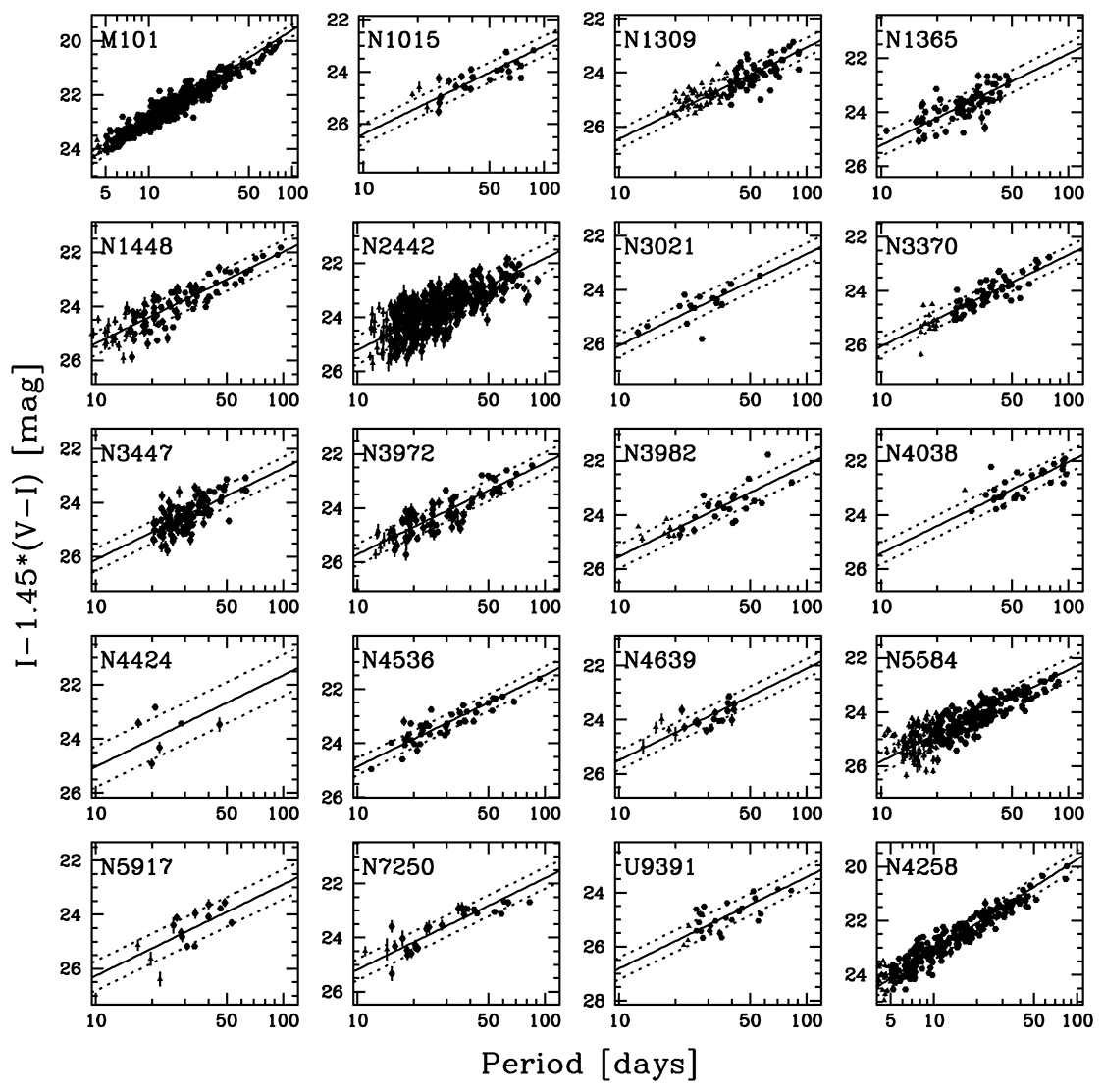}
\caption{W$_{I}$ PLRs for the Cepheids that passed all the selection criteria (except for minimum period cut, which was not applied to show the entire sample). Larger symbols denote variables located above the adopted period cuts. Solid lines indicate the $W_I$ PLR slope of $-3.38\pm0.02$~mag~dex$^{-1}$ derived in a global fit by R16, applicable to $P>10$~d. Dashed lines represent the observed $2\sigma$~dispersion of each PLR.}
\label{fig:megawespl}
\end{center}
\end{figure*}
 
We also required that the mean-light colors of the candidates fall within the expected range of the Cepheid instability strip, $0.5\!<V\!-\!I<\!1.5$~mag. We increased the upper limit to $V\!-\!I<\!2.0$~mag for {\tgal} and {\egal}, since they exhibited a larger amount of internal differential extinction and are subject to a greater mean foreground reddening owing to their relatively low Galactic latitude: $|b|\approx 13^\circ$ and $24^\circ$, respectively, with corresponding $E(V\!-\!I)$ of 0.202~mag and 0.268~mag \citep{schlafly11}. We note that the {\mgal} sample used by R16 accidentally included 5 variables with $1.5\!<\!V\!-\!I\!<\!1.65$~mag. These objects are included in our tables for completeness. Figure~\ref{fig:megacmd} displays the color-magnitude diagram of each galaxy. We plot the $2\sigma$ ``blue'' edge of the instability strip as derived from extinction-corrected colors of OGLE-III LMC Cepheids \citep{soszynski08}, shifted in distance modulus according to the values listed in Table~5 of R16. All our Cepheid samples exhibit colors that are consistent with this expected limit. We also plot in that figure the full instability strip shifted by distance modulus and by the mean color excess of each sample.

Given the lack of time-series information in {\V} and {\I} for a substantial fraction of the {\snh}, we did not apply any selection criteria based on the light-curve amplitude ratio between these two bands. Figure~\ref{fig:megamp} shows light-curve amplitudes in the primary band of each galaxy (either {\V} or {\W}) as a function of period.

We calculated ``Wesenheit'' magnitudes \citep{madore82}, which correct for the effects of extinction and the nonzero temperature width of the instability strip, with the formulation $W_{I}=I-{\cal{R}}_I\times(V\!-\!I)$. Using ground-based $V$ and $I$ magnitudes of LMC Cepheids provided by the OGLE project \citep{udalski99,soszynski08,soszynski15} and ${\cal{R}}_{I}=1.45$, we find a slope of $-3.309\pm{0.024}$~mag~dex$^{-1}$. We clipped $3\sigma$ outliers by sequentially removing the single most significant datum per iteration as advocated by \citet{kodric15}. In the case of {\wgal}, where the Cepheid candidate sample is very small and exhibits a large spread, we adopted a fixed range of $\pm1$~mag for the rejection. In the case of {\mgal}, the galaxy is significantly closer than the rest of the sample and the observations are comparatively deeper. Thus, we also detected Population~II pulsators that exhibit a well-separated parallel PLR about 2~mag fainter than the Population~I relation. We therefore removed 33 objects by applying an initial cut 1.5~mag below the center line of the Population~I relation. 

In the case of {\ngal}, we incorporate Cepheids from separate, unique surveys and thus are unable to apply the criteria in a universal manner. We apply the criteria to candidates when the necessary information is available to us. In particular, we fit the M06 objects from Tables~4 and 5 with the \citet{yoachim09} templates, which led to the inclusion of an additional 121 Cepheid candidates originally identified but rejected by M06, and we rejected 39 in their sample which failed to meet our requirements.

Figure~\ref{fig:megawespl} displays the $W_I$ PLRs of the ``Final I'' sample (column 12 of Table 4), augmented by 308 variables below the minimum period cuts that passed all other selection criteria, for a total of 2370 variables. We also plot in each panel the best-fit global slope of $-3.38\pm0.02$~mag~dex$^{-1}$ obtained by R16 for $P>10$~d. This slope was derived using fully calibrated {\it HST} {\V} and {\I}~magnitudes, and is therefore slightly different from the OGLE-based value quoted above.

Two additional restrictions are applied in the analysis detailed in our companion paper (R16): (1) variables with periods below a galaxy-dependent limiting value are not used, since the instability strip may not be fully sampled by our observations (see \S\ref{sec:pcomp}); and (2) objects must pass additional criteria related to their {\irh} photometry.  Table~\ref{tab:cut} documents the number of candidates remaining after each step in the selection process, and Table~\ref{tab:ceph} lists the properties of all the selected variables.

\section{Systematics}\label{sec:results}

\begin{figure}[t]
\begin{center}
\includegraphics[width=0.49\textwidth]{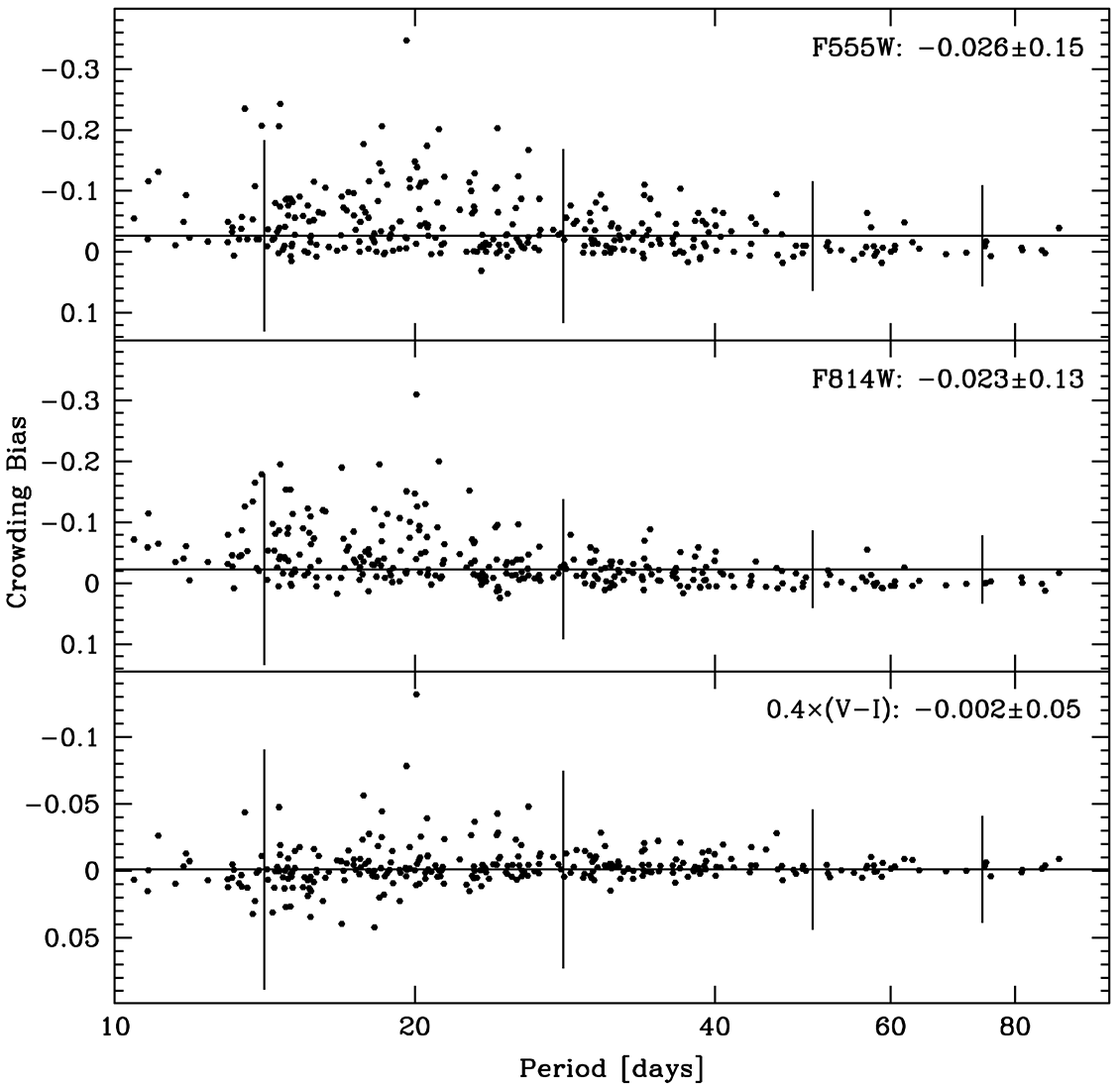}
\caption{Crowding corrections for Cepheids in {\tgal} that passed all selection criteria (except for minimum period cut, which was not applied to show the entire sample), obtained via artificial-star simulations. Solid lines indicate the mean values reported in Table~2 of R16. Error bars show the dispersion of magnitude offsets for multiple trials at the given period.}
\label{fig:ast}
\end{center}
\end{figure}

We now provide details of three systematic effects directly related to the discovery and characterization of Cepheids at optical wavelengths: photometric corrections owing to crowding, incomplete coverage of the instability strip at short periods, and determination of chemical abundances. A comprehensive analysis of other sources of systematic uncertainty that affect the determination of {\ho} is presented in R16.

\subsection{Crowding Corrections} \label{sec:crowd}

We define crowding as a systematic bias in photometric measurements arising from a high density of sources that in aggregate affect the statistical determination of the ``sky'' background and the PSF fit to an individual object. It is different from discrete blending, where some stars have a companion source (or in rarer cases, a few companion sources) that are too close to each other to be disentangled \citep{mochejska00,chavez12}. Blends can be removed by applying color cuts and iteratively rejecting outliers (see R16 for details) because the flux from the companion source(s) will make the Cepheid anomalously brighter and/or alter its color, rendering it an outlier. Crowding can be corrected through artificial-star simulations as described below.

We randomly added 10 stars with the same mean {\V} and {\I} magnitudes as a given Cepheid to the master image of each band, placing them within 30 pixels in radius relative to the location of each variable. Each artificial star was placed at the same location in both {\V} and {\I} to quantify the effect on the $V\!-\!I$ color. If an artificial star landed within 2.5~pixels of another source that was up to $3.5$~mag fainter, we flagged it as a blend; otherwise, it contributed to the crowding statistics. We repeated the procedure 20 times per galaxy to increase statistics.

We performed PSF photometry on the simulated images and compared the known input magnitude of the artificial stars to the measured values. We calculated a mean correction and dispersion for each Cepheid and used these values to derive a mean offset and dispersion for all objects in each galaxy. Figure~\ref{fig:ast} shows the result of this procedure for {\tgal}. The larger scatter seen at shorter periods is caused by the fainter nature of those Cepheids, whose photometric measurements are affected more strongly by variations in the the underlying stellar population. We adopted a single value for this correction for a given band and galaxy instead of a correction that varied with period (or logarithm of the period) because we did not find a statistically significant improvement from those approaches. Note that the crowding bias nearly cancels for the color combination used in the primary PLR of R16, a near-infrared Wesenheit magnitude. Table~2 in R16 presents further details of this procedure.

\subsection{PLR Incompleteness }\label{sec:pcomp}

Given the nonzero width of the instability strip and its projection into the period-luminosity plane, incomplete coverage below a certain period is a natural outcome of a magnitude-limited survey \citep{sandage88} if the photometry does not extend below the faintest Cepheid magnitudes. This leads to the preferential selection of brighter variables at shorter periods and, if unchecked, may result in a bias in the distance estimate.

\begin{figure*}[htbp]
\begin{center}
\includegraphics[width=\textwidth]{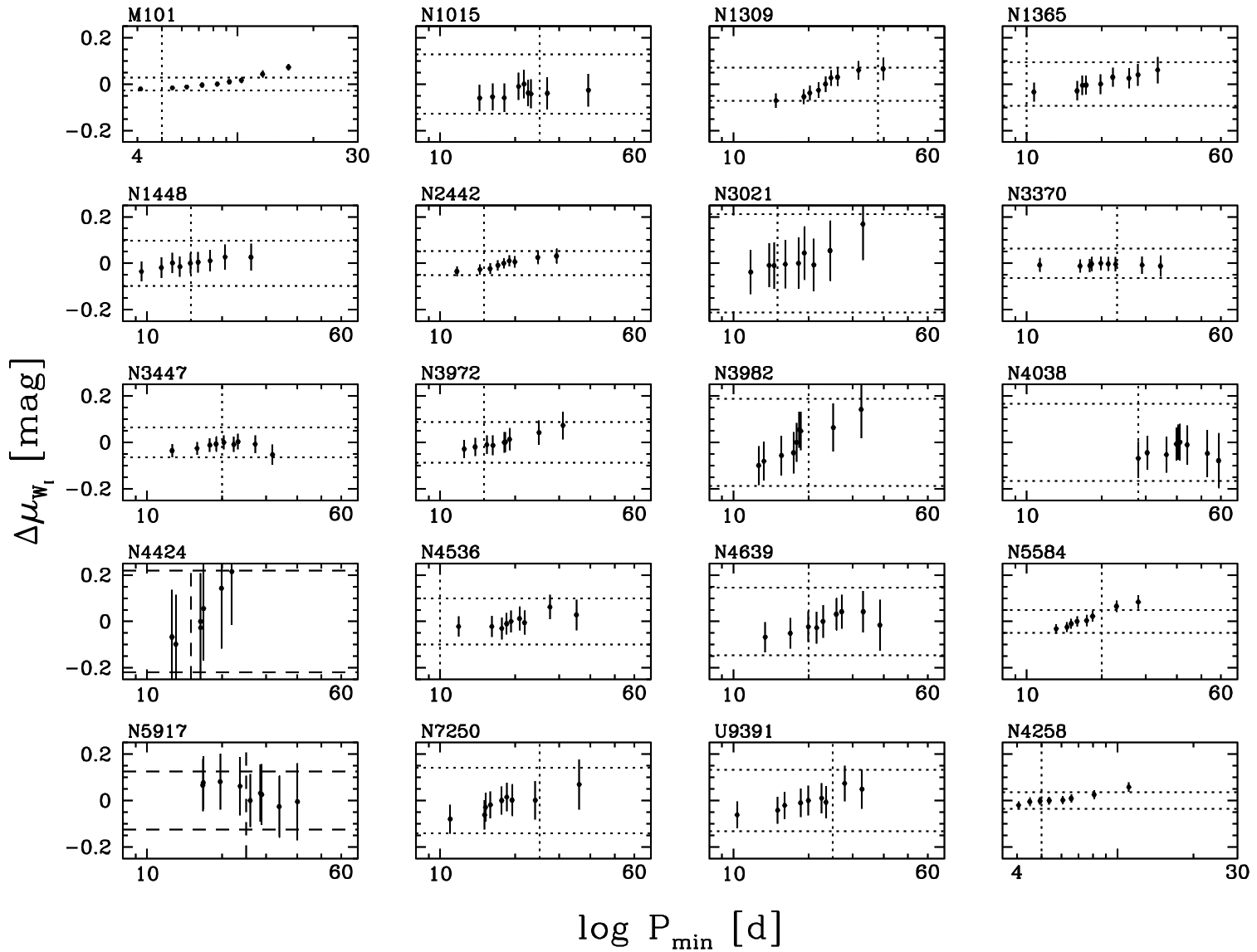}
\caption{Relative distance moduli and uncertainties in the mean as a function of minimum period for subsamples in each galaxy, based on fits to the $W_I$ PLRs. The symbols in each panel denote (from right to left) the results for subsamples containing the top 40, 55, 70, 75, 80, 85, 90, 95, and 100\% of the variables sorted by period. Values are expressed relative to the distance modulus obtained for the top 75\% of the sample (sorted by period). The horizontal dotted lines denote the $\pm 2\sigma$ range of values obtained by randomly subsampling 75\% of the variables in each galaxy (regardless of period); dashed lines are used to show the corresponding $1\sigma$ values for NGC$\,$4424 and NGC$\,$5917. While incompleteness below a given period is only sometimes apparent, conservative minimum period cuts are adopted and shown by the vertical dotted lines. In some cases these match the lowest-period Cepheid in a galaxy.}
\label{fig:pmin}
\end{center}
\end{figure*}

In previous iterations of this project (R09a and R11), we empirically derived completeness limits by calculating apparent distance moduli as a function of minimum period and identifying the values below which a bias in distance modulus became evident. While this varies slightly between galaxies, when examining {\ygal} we find a $S/N \sim10$ per epoch in {\V} at the period limit of 15~days. This $S/N$ value is a useful threshold below which we may expect some incompleteness due to several reasons described below. We carried out the same procedure for our new Cepheid samples, examining distance moduli in {\V}, {\W} (when available), {\I}, and $W_I$. Figure~\ref{fig:pmin} shows the result of this analysis for $W_{I}$. In some cases we see evidence for incompleteness below the aforementioned $S/N$ threshold, but in many we do not. Incompleteness may arise from inadequate phase coverage at periods below the lower limit of our optimal-spacing procedure (note the prominent aliasing features in the power spectrum of Figure~\ref{fig:alias} below 15~days). Photometric measurement errors, coupled to the use of Wesenheit magnitudes, can partially cancel the correlation of {\V} and {\I} residuals and dilute the effect of incompleteness below certain periods. Therefore, we adopted the minimum period cuts derived by R09 and R11 and applied conservative minimum period cuts to the new \snh\ based on their relative SN-based distances and the expected S/N for Cepheids of a given period based on exposure-time calculators. Table~\ref{tab:perlimit} lists the adopted values for all galaxies.

\begin{deluxetable}{lrrllrr}
\tablecaption{Minimum Period Cuts for Period-Luminosity Relations\label{tab:perlimit}}
\tablewidth{0.47\textwidth}
\tablehead{\multicolumn{1}{l}{Galaxy} & \multicolumn{1}{r}{$P_{\min}$} & \multicolumn{1}{r}{Note} & & \multicolumn{1}{l}{Galaxy} & \multicolumn{1}{r}{$P_{\min}$} & \multicolumn{1}{r}{Note}}
\startdata
\mgal &  5 & N  & \hspace{0.05\textwidth} & \rgal & 20 & VI \\
\fgal & 25 & W  & \hspace{0.05\textwidth} & \agal & 28 & VI \\
\xgal & 38 & VI & \hspace{0.05\textwidth} & \ngal &  5 & N  \\
\kgal & 10 & WF & \hspace{0.05\textwidth} & \wgal & 15 & W  \\
\ggal & 15 & W  & \hspace{0.05\textwidth} & \lgal & 10 & VI \\
\egal & 15 & W  & \hspace{0.05\textwidth} & \bgal & 20 & VI \\
\ygal & 15 & VI & \hspace{0.05\textwidth} & \tgal & 20 & VI \\
\zgal & 23 & VI & \hspace{0.05\textwidth} & \ogal & 25 & W  \\
\pgal & 20 & W  & \hspace{0.05\textwidth} & \dgal & 15 & W  \\
\qgal & 15 & W  & \hspace{0.05\textwidth} & \ugal & 25 & W    
\enddata
\tablecomments{N, nearby galaxy ($D\lesssim 7$~Mpc) with very deep photometry; VI, galaxy with time-series photometry in {\V} \& {\I} analyzed by R09a/R11, previously derived $P_{\rm min}$ adopted; W, galaxy with time-series photometry in {\W} only, conservative $P_{\rm min}$ adopted; WF, galaxy with previous WFPC2 photometry in {\V} and {\I} and deeper WFC3 master frames in those bands, conservative $P_{\rm min}$ adopted.\\ \\ \\}
\end{deluxetable}
 
\subsection{Chemical Abundance}\label{sec:chem}

The effect of different chemical abundances on Cepheid luminosities and colors, and therefore on distances derived by adopting ``universal'' PLRs, is a topic of intense investigation on both the observational and theoretical fronts (\citealt{gould94, macri06, romaniello08, bono10, freedman11, shappee11, pejcha12, kodric13, fausnaugh15}). While the effect is expected to be reduced and perhaps negligible at 1.6~$\mu$m, the analysis presented by R16 still solves for a ``metallicity dependence'' as a nuisance parameter and therefore requires an estimate of the chemical abundance of each Cepheid in the sample. As in previous iterations of this project, this is estimated from the metallicity gradient across each galaxy measured via emission-line spectroscopy of \ion{H}{2} regions (see \S2.5 of \citealt{riess05} and \S3 of R09a). Observations were carried out using Keck I/LRIS \citep{oke95} and supplemented by literature data where necessary.

\begin{deluxetable}{ccccccc}
\tabletypesize{\scriptsize}
\tablecaption{\ion{H}{2} Region Properties \label{tab:h2}}
\tablewidth{0pc}
\tablehead{\colhead{Galaxy} & \colhead{Name} & \colhead{$\alpha$} & \colhead{$\delta$} & \colhead{$r$} &\colhead{$Z$} & \colhead{$\sigma$} \\ \colhead{} & \colhead{} & \multicolumn{2}{c}{(J2000)} & \colhead{[kpc]} & \multicolumn{2}{c}{[dex]} }
\startdata
\fgal & H11 & 02:38:09.60 & -01:18:52.06  &    6.45 & 9.037  & 0.112  \\
\fgal & H12 & 02:38:10.66 & -01:18:32.66  &   11.93 & 9.139  & 0.065  \\
\fgal & H03 & 02:38:12.79 & -01:18:24.35  &   16.59 & 8.904  & 0.215  \\
\fgal & H09 & 02:38:11.78 & -01:18:28.13  &   14.17 & 8.864  & 0.561  
\enddata
\tablecomments{$Z=12+\log[{\rm O/H}]$. $r$: deprojected galactocentric radius using the distances from Table~5 of R16. Names starting with ``N'' indicate data observed after analysis and not utilized in the metallicity-gradient fits in Table~\ref{tab:metal}, but listed here for completeness. This table is available in its entirety in a machine-readable form in the online journal. A portion is shown here for guidance regarding its form and content.}  
\end{deluxetable}

\begin{deluxetable}{lrrrrrr}
\tabletypesize{\scriptsize}
\tablecaption{Galaxy Metallicity Gradients ($12 + \log$[O/H]) \label{tab:metal}}
\tablewidth{0pc}
\tablehead{\colhead{Galaxy} & \colhead{$Z$ ($r\!=\!3$~kpc)} & \colhead{$\sigma$} & \colhead{$dZ/dr$} & \colhead{$\sigma$} & \colhead{RMS} & \colhead{Note}\\
           \colhead{}       & \multicolumn{2}{c}{[dex]}                             & \multicolumn{2}{c}{[dex~kpc$^{-1}$]}           & \colhead{[dex]}      & \colhead{}}
\startdata
\mgal & 9.204 & 0.060 & -0.032 & 0.003 & 0.09 & a,f \\ 
\fgal & 9.423 & 0.178 & -0.050 & 0.020 & 0.18 &   f \\ 
\xgal & 9.075 & 0.057 & -0.072 & 0.010 & 0.08 &     \\ 
\kgal & 9.343 & 0.051 & -0.046 & 0.004 & 0.15 &     \\ 
\ggal & 9.083 & 0.063 & -0.030 & 0.004 & 0.15 &     \\ 
\egal & 9.315 & 0.168 & -0.050 & 0.020 & \nd  & b,f \\ 
\ygal & 9.154 & 0.082 & -0.121 & 0.027 & 0.08 &     \\ 
\zgal & 9.081 & 0.043 & -0.070 & 0.008 & 0.06 &     \\ 
\pgal & 8.810 & 0.178 & -0.050 & 0.020 & 0.23 &   f \\ 
\qgal & 9.232 & 0.115 & -0.050 & 0.020 & 0.12 &   f \\ 
\rgal & 9.042 & 0.051 & -0.117 & 0.019 & 0.09 &     \\ 
\agal & 9.113 & 0.046 & -0.041 & 0.009 & 0.14 &     \\ 
\ngal & 8.981 & 0.021 & -0.018 & 0.002 & 0.12 &     \\ 
\wgal & 9.000 & 0.081 & -0.050 & 0.020 & 0.08 &   f \\ 
\lgal & 9.071 & 0.033 & -0.036 & 0.004 & 0.06 &     \\ 
\bgal & 9.072 & 0.054 & -0.089 & 0.010 & 0.13 &     \\ 
\tgal & 8.968 & 0.040 & -0.067 & 0.007 & 0.06 &     \\ 
\ogal & 8.577 & 0.034 &    \nd & \nd   & 0.03 &   f \\ 
\dgal & 8.605 & 0.033 &    \nd & \nd   & 0.03 &   f \\ 
\ugal & 8.936 & 0.167 & -0.050 & 0.020 & 0.17 &   f    
\enddata
\tablecomments{$Z=12+\log[{\rm O/H}]$. $r$: deprojected galactocentric radius using the distances from Table~5 of R16. a: from \cite{kennicutt03}; b: from \cite{bajaja99}; f: gradients fixed to given values.}
\end{deluxetable}

\begin{figure}[t]
\begin{center}
\includegraphics[width=0.49\textwidth]{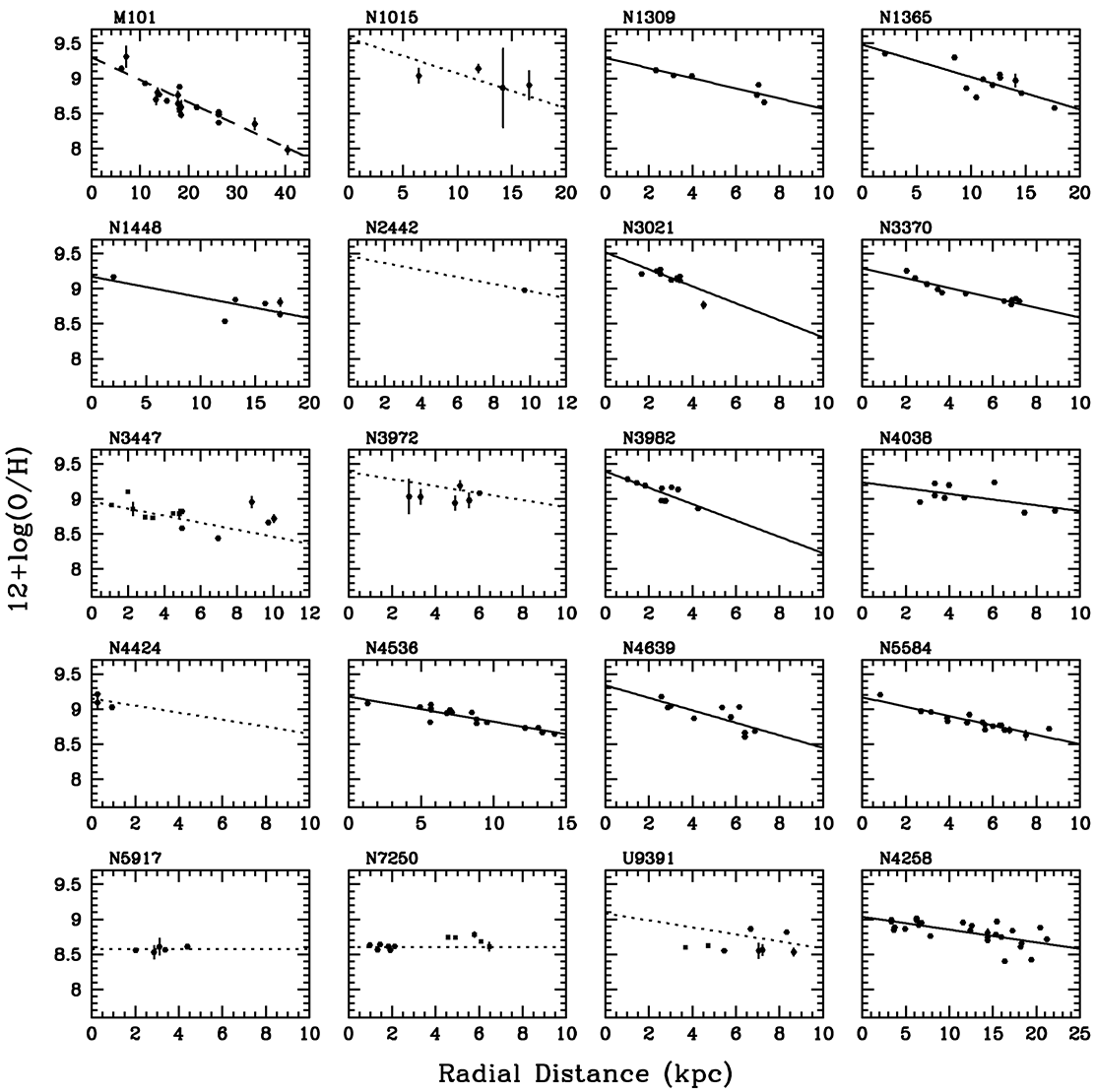}
\caption{Abundance values as a function of deprojected galactocentric radius. Solid lines indicate gradients obtained by fitting the data, while dotted lines indicate fixed values. The panel for {\mgal} shows the gradient derived by \citet{kennicutt03}, offset to the abundance scale of \citet{zaritsky94}. Square points represent data taken after the analysis was concluded and are shown for completeness.}
\label{fig:metal}
\end{center}
\end{figure}

We derived [O/H] abundances of individual \ion{H}{2} regions via the R$_{23}$ method as calibrated by \citet{zaritsky94} and present these measurements in Table~\ref{tab:h2}. We solved for a gradient as a function of deprojected galactocentric distance for galaxies with at least 6 measurements and low dispersion. In the case of galaxies that did not meet these criteria, we adopted a fixed gradient based on the mean of the aforementioned subsample. We were unable to obtain any observations of {\egal} and therefore calculated consistent abundances using the [\ion{O}{2}]/H$\beta$ and [\ion{O}{3}]/H${\beta}$ ratios published by \citet{bajaja99}. The use of a fixed global gradient led to implausibly low values for a small number of Cepheids in {\fgal} at large galactocentric radii, beyond the last \ion{H}{2} region. We set the abundances of those few variables to $12+\log({\rm O/H})=8.4$~dex, as the lowest value a Cepheid might realistically have in the disk of a large spiral \citep{bresolin12, sanchez16} in accordance with the minima seen in the other hosts. We solved for only a mean value in the case of intrinsically small galaxies where no spiral structure and \ion{H}{2} gradient is expected ({\ogal} and {\dgal}). Lastly, in the case of {\mgal} we adopted the metallicity values given in Equation~6 of \citet{kennicutt03}, but applied an overall offset to bring them into the system of \cite{zaritsky94}. Table~\ref{tab:metal} presents our findings, which are displayed in Figure~\ref{fig:metal}.

\citet{bresolin11} used an alternate calibration method based on the electron temperature ($T_e$) of nebular oxygen abundances which, as discussed by R11, consistently measures a shallower gradient compared to the \citet{zaritsky94} calibration. R16 used the metallicity of Cepheids derived from the measured gradients as a parameter and presented variants of {\ho} based on both calibrations as well as one independent of metallicity, allowing an examination of the changes in {\ho} due to the metallicity and these calibration methods. However, we emphasize that R16 find no metallicity dependence in the infrared analysis.

\section{Summary} \label{sec:sum}

We presented the result of a homogeneous search for Cepheids using {\it HST} at optical wavelengths in 19 {\snh} and {\ngal}, one of the anchors for the extragalactic distance scale. Our efforts yielded a sample of {\ncep} variables, the largest to date outside of the Local Group. We discussed our methodology for data processing, photometry, variability search, and identification of Cepheids, as well as systematic corrections required to enable a determination of {\ho} in our companion publication \citep{riess16}.

\acknowledgments

This project was funded by STScI through {\it HST} grants associated with the following programs: GO-12880, GO-13646, and GO-13647. S.L.H., L.M.M., and W.Y.~acknowledge financial support by the Mitchell Institute for Fundamental Physics \& Astronomy at Texas A\&M University. R.J.F.~gratefully acknowledges support from NASA grant 14-WPS14-0048, NSF grant AST-1518052, and the Alfred P.~Sloan Foundation. A.V.F.~has received generous financial assistance from the Christopher R.~Redlich Fund, the TABASGO Foundation, Gary \& Cynthia Bengier, and NSF grant AST-1211916. The work of A.V.F.~was completed at the Aspen Center for Physics, which is supported by NSF grant PHY-1066293; he thanks the Center for its hospitality during the black holes workshop in June and July 2016. Some of the data presented herein were obtained at the W.M.~Keck Observatory, which is operated as a scientific partnership among the California Institute of Technology, the University of California, and NASA; the observatory was made possible by the generous financial support of the W.M.~Keck Foundation. We are grateful to Melissa L.~Graham for assistance with the Keck observations and the Keck staff for their expert help. J.M.S.~is supported by an NSF Astronomy and Astrophysics Postdoctoral Fellowship under award AST-1302771. A.G.~and R.A.~acknowledge support from the Swedish National Space Board.

\bibliography{mega}{}
\bibliographystyle{apj}
\end{document}